\documentstyle[11pt]{article}
\normalbaselines

\newtheorem{thm}{Theorem}[section]
\newtheorem{lem}{Lemma}[section]
\newtheorem{prop}{Proposition}[section]

\begin{document}

\def\bea*{\begin{eqnarray*}}
\def\eea*{\end{eqnarray*}}
\def\ba{\begin{array}}
\def\ea{\end{array}}
\count1=1
\def\be{\ifnum \count1=0 $$ \else \begin{equation}\fi}
\def\ee{\ifnum\count1=0 $$ \else \end{equation}\fi}
\def\ele(#1){\ifnum\count1=0 \eqno({\bf #1}) $$ \else \label{#1}\end{equation}\fi}
\def\req(#1){\ifnum\count1=0 {\bf #1}\else \ref{#1}\fi}
\def\bea(#1){\ifnum \count1=0   $$ \begin{array}{#1}
\else \begin{equation} \begin{array}{#1} \fi}
\def\eea{\ifnum \count1=0 \end{array} $$
\else  \end{array}\end{equation}\fi}
\def\elea(#1){\ifnum \count1=0 \end{array}\label{#1}\eqno({\bf #1}) $$
\else\end{array}\label{#1}\end{equation}\fi}
\def\cit(#1){
\ifnum\count1=0 {\bf #1} \cite{#1} \else 
\cite{#1}\fi}
\def\bibit(#1){\ifnum\count1=0 \bibitem{#1} [#1    ] \else \bibitem{#1}\fi}
\def\ds{\displaystyle}
\def\hb{\hfill\break}
\def\comment#1{\hb {***** {\em #1} *****}\hb }

\newcommand{\TZ}{\hbox{\bf T}}
\newcommand{\MZ}{\hbox{\bf M}}
\newcommand{\ZZ}{\hbox{\bf Z}}
\newcommand{\NZ}{\hbox{\bf N}}
\newcommand{\RZ}{\hbox{\bf R}}
\newcommand{\CZ}{\,\hbox{\bf C}}
\newcommand{\PZ}{\hbox{\bf P}}
\newcommand{\QZ}{\hbox{\rm eight}}
\newcommand{\HZ}{\hbox{\bf H}}
\newcommand{\EZ}{\hbox{\bf E}}
\newcommand{\GZ}{\,\hbox{\bf G}}

\font\germ=eufm10
\def\goth#1{\hbox{\germ #1}}
\vbox{\vspace{38mm}}

\begin{center}
{\LARGE \bf Duality and Symmetry in  Chiral Potts Model  } \\[10 mm] 
Shi-shyr Roan \\
{\it Institute of Mathematics \\
Academia Sinica \\  Taipei , Taiwan \\
(email: maroan@gate.sinica.edu.tw ) } \\[25mm]
\end{center}

\begin{abstract}
We discover an Ising-type duality in the general $N$-state chiral Potts model, which is the Kramers-Wannier duality of planar Ising model when $N=2$. This duality relates the spectrum and eigenvectors of one chiral Potts model at a low temperature (of small $k'$) to those of another chiral Potts model at a high temperature (of $k'^{-1}$). The $\tau^{(2)}$-model and chiral Potts model on the dual lattice are established alongside the dual chiral Potts models. With the aid of this duality relation, we exact a precise relationship between the Onsager-algebra symmetry of a homogeneous superintegrable chiral Potts model and the $sl_2$-loop-algebra symmetry of its associated spin-$\frac{N-1}{2}$ XXZ chain through the identification of their eigenstates.  
\end{abstract}
\par \vspace{5mm} \noindent
{\rm 2008 PACS}:  05.50.+q, 02.20.Uw, 75.10Pq \par \noindent

\setcounter{section}{0}
\section{Introduction}
\setcounter{equation}{0}
The Kramers-Wannier duality is a reflective symmetry in statistical physics, which relates a two-dimensional square-lattice Ising model at a low temperature to another Ising model at a high temperature \cite{KW}. In this paper, we have found a Ising-type duality in the $N$-state chiral Potts model (CPM), which is the Kramers-Wannier duality of the usual Ising model when $N=2$. In the field of solvable statistical models, the $N$-state CPM has proved important and particularly challenging, due to the fact that for $N=2$ it reduces to Ising model, the free energy of which was calculated by Onsager in 1944. The model was originally formulated as an $N$-state one-dimensional Hamiltonian \cite{GR, HKN}, then as a two-dimensional classical lattice model in statistical mechanics \cite{AMPTY,BPA, MPTS}. The free energy was first obtained for infinity lattice using the properties of the free energy and its derivatives \cite{B88}. Then in 1990 the functional relations of \cite{BBP, BazS} were used to calculate the free energy more explicitly as a double integral \cite{B90, B91}. The interactions of CPM are defined by (local) Boltzmann weights depending on a temperature-like parameter $k'$, which is small at low temperatures and large at high temperatures. The system displays ferromagnetic order with a critical temperature below which the boundary conditions are relevant even for an infinitely large lattice. The order parameter has recently been proved by Baxter \cite{B05a, B05b}. Furthermore, results in \cite{AMP, B89} about the free energy in the superintegrable CPM  have strongly suggested the existence of duality in the theory of chiral Potts model parallel to the Kramers-Wannier duality in Ising model. In the present paper we will show that it is indeed the case. Here, we study the general inhomogeneous CPM of a finite size $L$ with a (skewed) boundary condition $r ~ (\in \ZZ_N)$. It is known that the chiral Potts transfer matrix, $T$ or $\widehat{T}$, with rapidities in $k'$-curve ${\goth W}$, carries a quantum number of $\ZZ_N$-charge $Q$,  (see (\req(cpmC)),(\req(Weig)) and (\req(ThatT)) in this paper). 
The duality of CPM relates two chiral Potts transfer matrices, $T$ and $T^*$, with the same eigenvalue spectrum, where 
$T$ is one over a $k'$-curve ${\goth W}$ in the $Q$-sector with the boundary condition $r$, and $T^*$ is another one over the $k'^{-1}$-curve ${\goth W}^*$ in the $Q^*$-sector with the boundary condition $r^*$, when the charge and boundary condition are interchanged, $(Q^*, r^*) = (r, Q)$. The duality is established upon the correspondence of rapidity curves, ${\goth W}$  and ${\goth W}^*$, about the dual Boltzmann weights, and a similar isomorphism of $(r, Q)$- and $(r^*, Q^*)$-quantum spaces about "ordered- and disordered-fields" (Theorem \ref{thm:dualCP} in the content). Indeed, the dual Boltzmann weights are connected by the relation of Fourier transform. Furthermore, under the dual correspondence of rapidities and quantum spaces, the $\tau^{(2)}$-models associated to two dual CPM are equally identified. In a special superintegrable case, the equal partition functions of the dual chiral Potts models  are in agreement with the duality discussion of Baxter in \cite{B89} where the vertical-interfacial-tension was computed. Similarly, we can form the general chiral Potts model over the dual lattice, as well as that for the face $\tau^{(2)}$-model, by using the underlying duality symmetry. In the homogeneous superintegrable case, there are two types of degeneracy symmetries about $\tau^{(2)}$-states in the study of CPM. One is the Onsager-algebra symmetry derived from the chiral-Potts-$\ZZ_N$-spin Hamiltonian, the other is the $sl_2$-loop-algebra symmetry induced from a twisted spin-$\frac{N-1}{2}$ XXZ chain which is equivalent to the $\tau^{(2)}$-model \cite{NiD, R05o, R06F}. We observe that the Onsager-algebra symmetries of two superintegrable $\tau^{(2)}$-models intertwine under the dual correspondence of rapidities and quantum spaces in the duality of CPM.  With the aid of this duality, all the $\tau^{(2)}$-degeneracy symmetries are unified in a common underlying  $(\otimes sl_2)$-structure  at $k'=0, \infty,$ for the eigenspace.

This paper is organized as follows. In section \ref{sec:CPM}, we recall some basic facts  in $\tau^{(2)}$ model and CPM. For the purpose of this paper, the results are formulated in the most general case, the inhomogeneous CPM with a skewed boundary condition. Much of the work in this section could be a paraphrase of those done before \cite{AMP, B90, B93, B94, BBP,BazS,Dav,GR, MR,R91,R06F,R0710,R0805}, but in a more general form suited for the discussion of this paper; some results will be simply stated with adequate citations at important places.  In section \ref{ssec.ICPb}, we first recall the main definitions in CPM, briefly discuss the functional relations and Bethe relation of the theory. Then we illustrate the results on the inhomogeneous superintegrable CPM, among which is a special periodic case appeared recently in \cite{ND08}. In section \ref{ssec.OACP}, we represent a detailed study of the homogeneous CPM with an arbitrary superintegrable vertical rapidity and the boundary condition, an extended superintegrable version for those in \cite{AMP,  B93, B94, GR}. By explicit formulas about chiral Potts transfer matrix and energy form of the $\ZZ_N$-spin Hamiltonian, the Onsager algebra symmetry and its induced $sl_2$-loop-algebra structure of the $\tau^{(2)}$-eigenspace are thoroughly discussed here. Section \ref{sec.DualCP} is devoted to the duality in CPM. First we discuss the $\tau^{(2)}$-duality in section \ref{ssec.Ftau}. By  studying $\tau^{(2)}$-face-model, we discover the duality of $\tau^{(2)}$-model through a correspondence of dual rapidity curves and quantum spaces. Based on the duality of rapidity curves for the dual Boltzmann weights, we verify the duality relation of CPM in section \ref{ssec.CPCP*}. In section \ref{ssec.CP*}, we incorporate the underlying duality symmetry of CPM into the formulation of chiral Potts model of the dual lattice, as well as that of the $\tau^{(2)}$-face-model. In section \ref{ssec.Ising}, we justify the CPM duality for $N=2$ in agreement with the usual Kramers-Wannier duality of Ising model \cite{Bax, KW}. Then in section \ref{ssec.DuOA}, we illustrate the consistency of the duality and quantum spin chain Hamiltonian in the homogeneous CPM. In particular, the 
Onsager-algebra symmetry of the dual homogeneous superintegrable models are identified under the duality transformation.
 In section \ref{sec:Uqsl}, we first recall the definition of a general inhomogeneous XXZ chain with the quantum group $U_q (sl_2)$ and an arbitrary skew boundary condition; then we derive the associated affine algebra $U_q (\widehat{sl}_2)$, and the root-of-unity-symmetry generators when $q^N=1$ for the XXZ chains with cyclic $U_q (sl_2)$-representation.  In subsection \ref{ssec.equiv}, using the argument in \cite{R0806} about the homogeneous case, we demonstrate the equivalent relation between the inhomogeneous $\tau^{(2)}$-models and XXZ chains with $U_q (sl_2)$-cyclic representation. In subsection \ref{ssec.symEq}, we study the relationship between all degeneracy symmetries of a homogeneous superintegrable $\tau^{(2)}$-model for odd $N$. By the equivalence between the $\tau^{(2)}, \tau_F^{(2)}$-model and homogeneous spin-$\frac{N-1}{2}$ XXZ chains which carry the $sl_2$-loop-algebra symmetry, we find that the $\tau^{(2)}$-Bethe-states can be identified with the highest or lowest weight vectors of the Onsager-algebra-Hamiltonian generators in a $\tau^{(2)}$-eigenspace. The canonical basis at $k'=0, \infty$ in the $\tau^{(2)}$-eigenspace provides a unified structure for both Onsager-algebra and $sl_2$-loop-algebra symmetry about the $\tau^{(2)}$-degeneracy. Finally we close in section \ref{sec.F} with a concluding remark.

Notation:  In this paper, we use the following standard notations. For a positive integer $N$ greater than one, 
$\CZ^N$ denotes the vector space of $N$-cyclic vectors with the canonical base 
$|\sigma \rangle, \sigma  \in \ZZ_N ~ (:= \ZZ/N\ZZ)$. We fix the $N$th root of unity $\omega = {\rm e}^{\frac{2 \pi {\rm i}}{N}}$, and  the Weyl $\CZ^N$-operators $X, Z$ :
$$
 X |\sigma  \rangle = | \sigma  +1 \rangle , ~ \ ~ Z |\sigma  \rangle = \omega^\sigma  |\sigma  \rangle ~ ~ \ ~ ~ (\sigma  \in \ZZ_N) ,
$$
satisfying $X^N=Z^N=1$ and the Weyl relation: $XZ= \omega^{-1}ZX$.  The Fourier basis $\{ \widehat{|k } \rangle \}$ of $\{ | \sigma  \rangle \}$ is defined by 
\be
\widehat{|k } \rangle  = \frac{1}{\sqrt{N}} \sum_{\sigma =0}^{N-1} \omega^{-k \sigma} |\sigma \rangle , ~ ~ ~ | \sigma  \rangle = \frac{1}{\sqrt{N}} \sum_{k =0}^{N-1} \omega^{\sigma k} \widehat{|k} \rangle , ~ ~ k, \sigma  \in \ZZ_N,
\ele(Fb) 
with the corresponding Weyl operators, $
\widehat{X} \widehat{|k} \rangle = \widehat{|k+1} \rangle$, $\widehat{Z} \widehat{|k} \rangle = \omega^k \widehat{|k} \rangle$. Then the following equality holds:
\be
(X, Z) = (\widehat{Z}, \widehat{X}^{-1}). 
\ele(XZF)

\section{$\tau^{(2)}$-model and Chiral Potts Model \label{sec:CPM}}
\setcounter{equation}{0}
The $L$-operator of $\tau^{(2)}$-model \cite{BBP, BazS, R0710, R0805}  is the two-by-two matrix expressed by Weyl $\CZ^N$-operators $X, Z$ or $\widehat{X}, \widehat{Z}$ in (\req(XZF)):
\bea(lll)
L ( t ) &= \left( \begin{array}{cc}
        1  -  t \frac{{\sf c}  }{\sf b' b} X ,  & (\frac{1}{\sf b }  -\omega   \frac{\sf a c }{\sf b' b} X) Z \\
       - t ( \frac{1}{\sf b'}  -  \frac{\sf a' c}{\sf b' b} X )Z^{-1}, & - t \frac{1}{\sf b' b} + \omega   \frac{\sf a' a c }{\sf b' b} X
\end{array} \right) \\
&= \left( \begin{array}{cc}
        1  -  t \frac{{\sf c}  }{\sf b' b} \widehat{Z} ,  & (\frac{1}{\sf b }  -\omega   \frac{\sf a c }{\sf b' b} \widehat{Z}) \widehat{X}^{-1} \\
       - t ( \frac{1}{\sf b'}  -  \frac{\sf a' c}{\sf b' b} \widehat{Z} )\widehat{X}, & - t \frac{1}{\sf b' b} + \omega   \frac{\sf a' a c }{\sf b' b} \widehat{Z}
\end{array} \right) \\ 
\elea(L)
with non-zero complex parameters  ${\sf a, b, a', b', c}$. It is known that the above $L$-operator satisfies the YB equation 
\be
R(t/t') (L (t) \bigotimes_{aux}1) ( 1
\bigotimes_{aux} L (t')) = (1
\bigotimes_{aux} L (t'))(L (t)
\bigotimes_{aux} 1) R(t/t') 
\ele(YBtau)
for the asymmetry six-vertex $R$-matrix
$$
R(t) = \left( \begin{array}{cccc}
        t \omega - 1  & 0 & 0 & 0 \\
        0 &t-1 & \omega  - 1 &  0 \\ 
        0 & t(\omega  - 1) &( t-1)\omega & 0 \\
     0 & 0 &0 & t \omega - 1    
\end{array} \right).
$$
Over the $\ell$th site of  a chain of size $L$, we consider the $L_\ell$-operator  with the parameter ${\sf a}_\ell, {\sf b}_\ell, {\sf a}'_\ell, {\sf b}'_\ell, {\sf c}_\ell$ in (\req(L)).  The monodromy matrix, 
\be
L_1 (t) L_2 (t) \cdots L_L (t) = \left( \begin{array}{cc}
        A(t)  & B(t) \\
        C(t) &  D(t)
\end{array} \right)  
\ele(Mont2) 
again satisfies the YB relation (\req(YBtau)). The $\tau^{(2)}$-model with
the (skewed) boundary condition 
\be
\sigma_{L+1} \equiv \sigma_1 - r \pmod{N} ,  ~ ~ ~ ( r \in \ZZ ),
\ele(sBy)
is the commuting family of operators defined by 
\be
\tau^{(2)}(t) = A(\omega t) + \omega^r D(\omega  t)  
\ele(tau2) 
which commute with the spin-shift operator $X (:= \prod_{\ell} X_\ell)$. The eigenvalues of $X$ will be denoted by $\omega^Q ~ (Q \in \ZZ_N)$. The general $\tau^{(2)}$-model (\req(tau2)) with arbitrary parameters $\{{\sf a}'_\ell, {\sf b}'_\ell, {\sf a}_\ell, {\sf b}_\ell, {\sf c}_\ell \}_\ell$ was studied by Baxter in \cite{B049}. In this paper, we consider only the $\tau^{(2)}$-model in CPM, with the parameters, 
\be
({\sf a}'_\ell,  {\sf b}'_\ell, {\sf a}_\ell, {\sf b}_\ell, {\sf c}_\ell)= ( x_{p'_\ell}, 
y_{p'_\ell}, x_{p_\ell}, y_{p_\ell}, \mu_{p'_\ell}\mu_{p_\ell}), 
\ele(pp'l)
where $ p_\ell, p'_\ell$ ($ 1 \leq \ell \leq L $ ) all lie in the {\it same} rapidity curve of CPM for a temperature-like parameter $k' ( \neq 0)$, i.e., a curve ${\goth W}= {\goth W}_{k'} (= {\goth W}_{k', k} ), {\goth W}_{\pm 1}, \overline{\goth W}_{\pm 1}$ consisting of elements  $(x, y, \mu) \in \CZ^3$ with the equation
\bea(ll)
{\goth W}_{k'}&: k x^N  = 1 -  k'\mu^{-N},   k  y^N  = 1 -  k'\mu^N, \ ( k'^2 \neq 1,   k^2 + k'^2 = 1 ) ; \\
{\goth W}_1  &:  x^N + y^N =  1, \ \mu^N=1 ;   \\
{\goth W}_{-1} &:  x^N = 1 + \mu^{-N}, \ ~   y^N = 1 + \mu^N ;   \\
\overline{\goth W}_{\pm 1} &: x^N + y^N = 0, \ \mu^N =  \pm 1  ~ ~ ~ {\rm respectively}
\elea(cpmC)
(see, e.g. \cite{AMP, BBP, R0710, R0805}\footnote{The curves ${\goth W}_1, {\goth W}_{-1}, \overline{\goth W}_{\pm 1}$ here correspond to ${\goth W}_1^{\prime \prime}, {\goth W}_1^{\prime},{\goth W}_1^{\prime \prime \prime}$ respectively in \cite{R0805} (2.9).} ). Note that for a given $k' \neq \pm 1$, ${\goth W}_{k', k}$ is isomorphic to ${\goth W}_{k', -k}$ via the transformation $(x, y, \mu) \mapsto ((-1)^\frac{1}{N}x, (-1)^\frac{1}{N}y, \mu)$. Hereafter we shall write ${\goth W}_{k'}$ to represent one of these two curves  if no confusion could arise; and we write the $\tau^{(2)}$-model (\req(tau2)) with parameters (\req(pp'l)) by 
\be
\tau^{(2)}(t) = \tau^{(2)}(t ; { \{p_\ell \}, \{ p'_\ell \} }) .
\ele(t2inh)
The spectral parameter $t$ of $\tau^{(2)}$ will be identified with $x_q y_q$ for a generic rapidity $q$ of a curve ${\goth W}$ in (\req(cpmC)):
$$
t ~ ( = t_q) = x_q y_q .
$$
Then  $x^N$ is related to $t^N$ by a quadratic relation, which defines a hyperelliptic curve $W= W_{k'}, W_{\pm 1}, \overline{W}_{\pm 1} $  of lower genus with the coordinates $(t, \lambda)$ (\cite{R0805} (2.13) (2.16)):
\bea(lll)
W_{k'} :  t^N = \frac{(1- k' \lambda  )( 1 - k' \lambda^\dagger) }{k^2 } , & 
W_{-1} :  t^N= (1 + \lambda )(1+ \lambda^\dagger ) , & ( \lambda := \mu^N , \lambda^\dagger = \lambda^{-1} ) ,  \\
W_1 : t^N = \lambda \lambda^\dagger ,   & ( \lambda := x^N, \lambda^\dagger = 1- \lambda ), & \\
\overline{W}_{\pm 1} : t^N =  \lambda \lambda^\dagger , &( \lambda := x^N, \lambda^\dagger = - \lambda ).   
\elea(hW)
In the case $\overline{W}_{\pm 1}$, only odd $N$ case will be considered  as the curve in even $N$ case consists of two rational irreducible components. As in \cite{B93} (3.11)-(3.13), \cite{R06F} Proposition 2.1, (2.31) and \cite{R0710} (2.25), one can construct $\tau^{(j)}$-matrices from the $L$-operator (\req(L)), with $\tau^{(0)}=0, \tau^{(1)}= I$ and $\tau^{(2)}$ in (\req(t2inh)), so that the fusion relation holds:
\bea(l)
\tau^{(2)}(\omega^{j-1} t) \tau^{(j)}(t) =  \omega^r X z( \omega^{j-1} t) \tau^{(j-1)}(t)  + \tau^{(j+1)}(t) , \ \ j \geq 1 ; \\
\tau^{(N+1)}(t) = \omega^r X z(t) \tau^{(N-1)}(\omega t) + u(t) I ,
\elea(fus)
where $z(t), u(t)= \alpha_q  + \overline{\alpha}_q$ are defined by 
\bea(l)
z (t)= \prod_{\ell=1}^L  \frac{\omega \mu_{p_\ell} \mu_{p'_\ell}(t_{p_\ell}-t)(t_{p'_\ell}-t)}{y_{p_\ell}^2 y_{p'_\ell}^2} , \\
\alpha_q = \prod_{\ell=1}^L  \frac{\mu^N (y_{p_\ell}^N-x^N) (y_{p'_\ell}^N-x^N)}{k' y_{p_\ell}^N y_{p'_\ell}^N} , ~ ~  \overline{\alpha}_q   = \prod_{\ell=1}^L \frac{\mu^{-N} (y_{p_\ell}^N-y^N) (y_{p'_\ell}^N-y^N)}{k' y_{p_\ell}^N y_{p'_\ell}^N}.
\elea(abF)

\subsection{Inhomogeneous chiral Potts model with a skewed boundary condition \label{ssec.ICPb}}
With the rapidities $p, q \in {\goth W}$ in (\req(cpmC)), 
the Boltzmann weights of CPM are defined by
\be
\frac{W_{p q}(\sigma)}{W_{p q}(0)}  = (\frac{\mu_p}{\mu_q})^\sigma \prod_{j=1}^\sigma
\frac{y_q-\omega^j x_p}{y_p- \omega^j x_q }  , \ ~ \
\frac{\overline{W}_{p q}(\sigma)}{\overline{W}_{p q}(0)}  = ( \mu_p\mu_q)^\sigma \prod_{j=1}^\sigma \frac{\omega x_p - \omega^j x_q }{ y_q- \omega^j y_p }, 
\ele(Weig)
which satisfy the star-triangle relation \cite{AMPT, AuP, BPA, FatZ,  MaS, MPTS}
\be
\sum_{\sigma=0}^{N-1} \overline{W}_{qr}(j' - \sigma) W_{pr}(j - \sigma) \overline{W}_{pq}(\sigma - j'')= R_{pqr} W_{pq}(j - j')\overline{W}_{pr}(j' - j'') W_{qr}(j -j'')    
\ele(TArel)
where $R_{pqr}= \frac{f_{pq}f_{qr}}{f_{pr}}$ with $f_{pq} =  (\frac{\overline{g}_p(q) }{g_p (q)})^{1/N}$, and 
\bea(lll)
g_p(q)&: =  \prod_{n=0}^{N-1} W_{pq}(n)& = (\frac{\mu_p}{\mu_q})^{(N-1)N/2} \prod_{j=1}^{N-1} (\frac{  x_p -\omega^j y_q }{  x_q- \omega^j y_p })^j, \\
\overline{g}_p (q)&: = {\rm det}_N(\overline{W}_{p q}(i-j))&= N^{N/2} {\rm e}^{{\rm i} \pi (N-1)(N-2)/12} \prod_{j=1}^{N-1} \frac{(t_p - \omega^j t_q)^j }{(x_p - \omega^j x_q)^j (y_p - \omega^j y_q)^j },
\elea(gg1)
(\cite{BBP} (2.44), \cite{R0805} (2.24)).
Without loss of generality, we set $W_{p,q}(0)= \overline{W}_{p,q}(0)=1$. The $N$-cyclic vectors defined by the Boltzmann weights (\req(Weig)) can also be expressed in terms of the Fourier bases: $
\sum_{\sigma =0}^{N-1} W_{pq}(\sigma) | \sigma \rangle = \sum_{k =0}^{N-1} W^{(f)}_{pq}(k) \widehat{|k} \rangle $, $ 
\sum_{\sigma =0}^{N-1} \overline{W}_{pq}(\sigma) | \sigma \rangle = \sum_{k =0}^{N-1} \overline{W}^{(f)}_{pq}(k) \widehat{|k} \rangle$. By \cite{BBP} (2.24), one finds
\bea(ll)
\overline{W}^{(f)}_{pq}(k)= \frac{1}{\sqrt{N}} \sum_{\sigma = 0}^{N-1} \omega^{k \sigma }\overline{W}_{pq}(\sigma) , & \frac{\overline{W}^{(f)}_{pq}(k)}{\overline{W}^{(f)}_{pq}(0)} = \prod_{j=1}^k \frac{y_q - \omega^j x_p \mu_p \mu_q }{y_p - \omega^j x_q \mu_p \mu_q} , \\
W^{(f)}_{pq}(k) = \frac{1}{\sqrt{N}} \sum_{\sigma = 0}^{N-1} \omega^{k \sigma }W_{pq}(\sigma) , & \frac{W^{(f)}_{pq}(k)}{W^{(f)}_{pq}(0)} = \prod_{j=1}^{N-k} \frac{\omega x_p \mu_p - \omega^j x_q  \mu_q }{y_q \mu_p - \omega^j y_p \mu_q}. 
\elea(WWf)
The chiral Potts transfer matrix of a size $L$ with the (skewed) boundary condition (\req(sBy)) and vertical rapidities $\{ p_\ell , p'_\ell \}_{\ell=1}^L $ are the $\stackrel{L}{\otimes} \CZ^N$-operators defined by (\cite{B93, BBP})
\bea(ll)
T (q)_{\{\sigma \}, \{\sigma'\}} &(= T (q ; { \{p_\ell \}, \{ p'_\ell \} })_{\{\sigma \}, \{\sigma'\}}) = \prod_{\ell =1}^L W_{p_\ell q}(\sigma_\ell - \sigma'_\ell ) \overline{W}_{p'_\ell q}(\sigma_{\ell+1} - \sigma'_\ell), \\
\widehat{T} (q)_{\{\sigma' \}, \{\sigma'' \}}&(=\widehat{T} (q ; { \{p_\ell \}, \{ p'_\ell \} } )_{\{\sigma' \}, \{\sigma''\}}) = \prod_{\ell =1}^L \overline{W}_{p_\ell q}(\sigma'_\ell - \sigma''_\ell) W_{p'_\ell q}(\sigma'_\ell - \sigma''_{\ell+1}) ,
\elea(ThatT)
which commute with the spin-shift operator $X$. Here $q$ is an arbitrary rapidity, $\sigma_\ell, \sigma'_\ell \in \ZZ_N$, and the periodic vertical rapidities, $p_{L+1}=p_1 , p'_{L+1}=p'_1$, are imposed. The star-triangle relation (\req(TArel)) yields the relation 
$$
T (q) \widehat{T} (r) = (\prod_{\ell =1}^L \frac{f_{p'_\ell q}f_{p_\ell r}}{f_{p_\ell q}f_{p'_\ell r}}) T (r) \widehat{T} (q) , \ \ \widehat{T} (q) T (r) = (\prod_{\ell =1}^L \frac{f_{p_\ell q}f_{p'_\ell r}}{f_{p'_\ell q}f_{p_\ell r}})  \widehat{T} (r) T (q) ,
$$
by which the following commutative relations hold for rapidities $q, r, q', r'$:
$$
\widehat{T} (q) T (r) \widehat{T} (q') T (r') = \widehat{T} (q') T (r') \widehat{T} (q) T (r), ~ ~ T (q) \widehat{T} (r) T (q') \widehat{T} (r') = T (q') \widehat{T} (r') T (q) \widehat{T} (r). 
$$
Hence the matrices $T (q), \widehat{T}(q)$ can be diagonalized by two invertible $q$-independent matrices $P_B, P_W$, i.e.,  $P_W^{-1} T (q) P_B, P_B^{-1} \widehat{T}(q)P_W$ are diagonal with the "eigenvalues" of $T, \widehat{T}$ as the diagonal entries (\cite{BBP} (2.32)-(2.34), (4.46), \cite{B93} (2.10)-(2.13)):
\be
\widehat{T}_{\rm diag} (q) =   T_{\rm diag} (q)(\prod_{\ell=1}^L \frac{f_{p_\ell q} }{f_{p'_\ell q}}) D
\ele(TTD)
where $D$ is $q$-independent diagonal matrix. In particular for the homogeneous case, there is one extra  symmetry:
\be
\widehat{T} (q)  = T (q) S_R  = S_R T (q) , ~ ~ {\rm when} ~ p_\ell = p'_\ell = p ~ ~ {\rm for ~ all} ~ \ell ,
\ele(homCP)
where $S_R$ is the spatial translation operator $(S_R)_{\sigma, \sigma'} = \prod_\ell \delta_{\sigma_{\ell-1} \sigma'_\ell}$, equivalently  $S_R | j_1, \ldots, j_L \rangle = | j_2,  \ldots, j_{L+1} \rangle$ or $\langle  j_1, \ldots, j_L |S_R = | j_0,  j_1, \ldots, j_{L-1} \rangle$, with eigenvalues $\omega^{-rQ/L}{\rm e}^{2 \pi l_R/L}~ (l_R \in \ZZ_L )$. Then $\{ T (q), \widehat{T} (q)\}_{q \in {\goth W}}$ form a commuting family, with $P_B=P_W$ and $D=S_R$ in (\req(TTD)). 

In the study of inhomogeneous chiral Potts model (\cite{B93}, \cite{BBP} page 842), there are functional relations between $T, \widehat{T}$ in (\req(ThatT)) and $\tau^{(2)}, \tau^{(j)}$ in (\req(fus)). The $\tau^{(2)}T$-relation is the relation between the $\tau^{(2)}$- and $T$-matrices (\cite{B93} (3.15), \cite{BBP} (4.31), \cite{R0805} (2.31)-(2.32)):  
\bea(l)
\tau^{(2)}(t_q) T (Uq) =  \varphi_q T( q) + \omega^r \overline{\varphi}_{Uq} X T (U^2 q) ,  \\
\tau^{(2)}(t_q) T(U'q) = \omega^r \varphi_q^\prime  X T(q) + \overline{\varphi}_{U'q}^\prime T(U'^2 q) , 
\elea(tauTU)
where the automorphism $U(x,y,\mu):=(\omega x,y,\mu)$, $ U'(x,y,\mu):=(x,\omega y,\mu)$, and the functions $\varphi_q , \overline{\varphi}_q , \varphi_q^\prime, \overline{\varphi}_q^\prime$ are defined by
$$
\begin{array}{ll}
\varphi_q  (= \varphi_{\{p_\ell \}, \{p'_\ell \}; q}) = \prod_{\ell} \frac{(t_{p'_\ell}- t_q)  (y_{p_\ell}-  \omega x_q)}{y_{p_\ell} y_{p'_\ell}(x_{p'_\ell}-  x_q)}, & \overline{\varphi}_q (=\overline{\varphi}_{\{p_\ell \}, \{p'_\ell \};q}) =
\prod_{\ell} \frac{\omega \mu_{p'_\ell} \mu_{p_\ell}(t_{p_\ell}- t_q)(x_{p'_\ell}- x_q) }{y_{p_\ell} y_{p'_\ell}(y_{p_\ell}- \omega x_q)}  , \\
\varphi_q^\prime  (= \varphi_{\{p_\ell \}, \{p'_\ell \};q}^\prime ) =  
\prod_{\ell} \frac{\omega \mu_{p_\ell} \mu_{p'_\ell}(t_{p'_\ell}- t_q)(x_{p_\ell}- y_q) }{y_{p_\ell} y_{p'_\ell}(y_{p'_\ell}- y_q)}, &  
\overline{\varphi}_q^\prime  (=\overline{\varphi}_{\{p_\ell \}, \{p'_\ell \}; q}^\prime ) = \prod_{\ell} \frac{(t_{p_\ell}- t_q)  (y_{p'_\ell}-  y_q)}{y_{p_\ell} y_{p'_\ell}(x_{p_\ell} -  y_q)}.
\end{array}
$$
Similarly, one has the $\tau^{(2)}T$-relation between $\tau^{(2)}$ and $\widehat{T}$ (\cite{R0805} (2.33)):
$$ 
\begin{array}{l}
\widehat{T}(U q) \tau^{(2)}(t_q)  =  \varphi_{\{p'_\ell \},\{p_\ell \}; q} \widehat{T}( q) + \omega^r \overline{\varphi}_{\{p'_\ell \}, \{p_\ell \}; Uq}  X \widehat{T}(U^2 q) , \\
\widehat{T}( U' q) \tau^{(2)}(t_q)  = \omega^r \varphi_{\{p'_\ell \}, \{p_\ell \}; q}^\prime  X \widehat{T}(q) + \overline{\varphi}_{\{p'_\ell \}, \{p_\ell \}; U'q}^\prime  \widehat{T}(U'^{2} y_q) .
\end{array}
$$
Using (\req(tauTU)) and (\req(fus)), one finds $\tau^{(j)}T$-relation (\cite{BBP} $(4.34)_{k=0}$,  \cite{R0805} (2.34)):
$$
\tau^{(j)}(t_q)= \sum_{m=0}^{j-1} \omega^{rm} \varphi_q \cdots \varphi_{U^{m-1}q}
\overline{\varphi}_{U^{m+1}q}\cdots \overline{\varphi}_{U^{j-1}q} T(q)T(U^m q)^{-1} T(U^j q) T(U^{m+1}q)^{-1}X^{j-m-1} .
$$
Other than the above $\tau^{(j)}T$- and $\tau^{(2)}T$-relations, there is the $T\hat{T}$-relation (\cite{B93} (3.1) \cite{BBP} (2.36) \cite{B02} (13), \cite{R0805} (4.7)):
\be
 \frac{T(q) \widehat{T}(y_q, \omega^j x_q, \mu_q^{-1})}{r(q) h_{j}(q) } = \tau^{(j)} (t_q) + \omega^{jr} \frac{z(t_q)z(\omega t_q) \cdots z(\omega^{j-1} t_q)}{\alpha_q } \tau^{(N-j)} (\omega^j t_q) X^j  
\ele(TT)
where   $z(t), \alpha_q$ are in (\req(abF)), and $r(q), h_{j}(q)$ are defined by  
$$
\begin{array}{ll}
r(q)= \prod_\ell \frac{N(x_{p'_\ell}-x_q)(y_{p'_\ell}-y_q)(t_{p'_\ell}^N -t_q^N)}{(x_{p'_\ell}^N- x_q^N) (y_{p'_\ell}^N -y_q^N)(t_{p'_\ell}-t_q)},& h_{j}(q)=\prod_\ell  \prod_{m=1}^{j-1} \frac{y_{p_\ell} y_{p'_\ell} (x_{p'_\ell}- \omega^m x_q)}{(y_{p_\ell}-\omega^m x_q)(t_{p'_\ell}-\omega^m t_q)}.
\end{array}
$$
Then the functional relation of $T, \widehat{T}$ (\cite{BBP}(4.40)) follows from (\req(TT)) and the $\tau^{(j)}T$-relation.
One can solve the eigenvalue problem of CPM using the whole set of functional relations \cite{B90, MR, R0805}. First, we need to solve $\tau^{(2)}$-eigenvalues satisfying the following Bethe relation (\cite{R0805} (3.7) (3.10)), a parallel version of the $\tau^{(2)}T$-relation (\req(tauTU)), 
\bea(ll)
\tau^{(2)}(t) &= \omega^{-P_a} h^+(t) \frac{ F (t)}{F( \omega t)} + \omega^{Q+P_a+r} h^-(\omega t)\frac{F(\omega^2 t)}{F( \omega t)} \\ 
&= \omega^{Q-P_b+r}  h'^-(t) \frac{ F' (t)}{F'( \omega t)} + \omega^{P_b} h'^+ (\omega t)\frac{ F'(\omega^2 t)}{F'( \omega t)}
\elea(t2F)
through the Bethe polynomial $t^{P_a} F(t) = t^{P_b} F'(t)$ with $F (t) = \prod_{j=1}^J (1+ \omega v_j t)$ and $F' (t) = \prod_{j=1}^{J'} (1+ \omega v'_j t)$. Here the $t$-functions $h^\pm (t)$ or $h'^\mp (t)$ are obtained through the Wiener-Hopf splitting of $\alpha_q, \overline{\alpha}_q$ in (\req(abF)) (for the details, see \cite{R0805} section (3.2)). The regular-function condition of $\tau^{(2)}(t)$ gives rise to the Bethe equation of  $v_j$s  or $v'_j$s (see, formulas (3.8),(3.11) in \cite{R0805}).
By (\req(fus)), one then express the functions $\tau^{(j)}(t) ~ (j \geq 2 )$ in terms of the Bethe solution $F(t), F'(t)$ (\cite{R0805} (3.9) (3.12)):
$$
\begin{array}{ll}
\tau^{(j)}(t)&= \omega^{(j-1)(Q+P_a+r)} F(t ) F(\omega^j t ) \sum_{k=0}^{j-1} \frac{h^+(t) \cdots h^+(\omega^{k-1} t) 
h^-( \omega^{k+1}t) \cdots h^-(\omega^{j-1} t)\omega^{-n(Q+2P_a+r)} }{ 
 F(\omega^k t)  F(\omega^{k+1}t)}  \\
&= \omega^{(j-1)P_b} F'(t ) F'(\omega^j t ) \sum_{k=0}^{j-1} \frac{h'^-(t) \cdots h'^-(\omega^{k-1} t) h'^+( \omega^{k+1}t) \cdots h'^+(\omega^{j-1} t)\omega^{k(Q-2P_b+r )} }{ F'(\omega^k t)  F'(\omega^{k+1}t)}. 
\end{array}
$$
Using the above $\tau^{(N)}(t)$-expressions and the $T\hat{T}$-relation (\req(TT)) for $j=N$, one can derive the expression of eigenvalues of $T (q), \widehat{T}(q)$ through the functional relation method (for the details, see \cite{R0805} section 4)\footnote{The argument and formulas about Bethe equation of  $\tau^{(2)}$-model and eigenvalues of chiral Potts transfer matrix with alternating rapidities in \cite{R0805} are all valid in the inhomogeneous and skewed-boundary condition case after a suitable modification of scalar coefficients as described in \cite{B93} and \cite{BBP} page 842. }. 

We now describe the formulas in the superintegrable case (\cite{AMP, B93, B94}, \cite{R0805} section 4.3). 
In this paper, by the superintegrable\footnote{The superintegrable condition here is slightly different from the alternating superintegrable case in \cite{R0805} (4.30), where the integers ${\sf m}, {\sf m}'$ are assumed to be equal in the discussion of this paper. } inhomogeneous CPM, we mean the vertical rapidities $\{ p_\ell , p'_\ell \}_{\ell=1}^L $ satisfy the relation,
\be  
x_{p_\ell} = \omega^m  y_{p'_\ell}, ~ ~    x_{p'_\ell} = \omega^m  y_{p_\ell}, ~ ~ ~  \mu_{p_\ell} \mu_{p'_\ell}= \omega^n, 
\ele(supdef)
for some integers $m, n$, which is equivalent to $t_{p_\ell} = t_{p'_\ell} = \omega^m y_{p_\ell} y_{p'_\ell}$,  $\mu_{p_\ell} \mu_{p'_\ell}= \omega^n$. Denote 
$$
h(t) := \prod_{\ell=1}^L (1- \frac{t}{t_{p_\ell}} ).
$$
The functions $h^{\pm} (t), h'^{\pm} (t)$ in (\req(t2F)) are given by $h^{+} (t)= h'^+ (t) = h(t)$, $h^{-} (t)= h'^- (t) = \omega^{(1+2m +n)L} h(t)$ with the $\tau^{(2)}$ and $\tau^{(N)}$-eigenvalues expressed by (\cite{R0805} (3.7)-(3.11))
\bea(l)
\tau^{(2)}(t) = \omega^{-P_a} h (t) \frac{ F (t)}{F( \omega t)} + \omega^{P_b} h (\omega t)\frac{F(\omega^2 t)}{F( \omega t)} ,  \\
\frac{\tau^{(N)}(t)}{F(t)^2} = \omega^{-P_b} \sum_{k=0}^{N-1} \frac{h(t)\cdots h( \omega^{k-1}t) h( \omega^{k+1}t)\cdots h( \omega^{N-1}t)\omega^{-k(P_a+P_b)}}{F(\omega^k t) F( \omega^{k+1} t) } 
\elea(tausp)
where the polynomial $F (t) = \prod_{j=1}^J (1+ \omega v_j t)$ satisfies the Bethe equation (\cite{R0805} (4.32)):
\be
 \prod_{\ell=1}^L \frac{(t_{p_\ell} v_i + \omega^{-1}  )}{(t_{p_\ell} v_i + \omega^{-2}  )}  = - \omega^{-P_a- P_b} \prod_{j=1}^J \frac{v_i -  \omega^{-1}   v_j }{ v_i -\omega  v_j } , \ \ i= 1, \ldots, J.
\ele(Besupin)
Note that the right-hand side of the above equation is equal to $\frac{h (-  \omega^{-1} v_i^{-1})}{h (-  \omega^{-2} v_i^{-1} )}$. 
Here $P_a, P_b$ are  are integers satisfying the relations (\cite{R0805} (4.36) (4.37))\footnote{A misprint occurred in the last formula in \cite{R0805} (4.37) where $J+P_b \equiv ({\tt m} +2 {\tt m}')L$ should be $J+P_b \equiv {\tt m} L$. }
\bea(ll)
0 \leq P_a + P_b \leq N-1, &P_b - P_a \equiv Q +r +(1+2m +n)L \pmod{N} ; \\
P_a \equiv 0 ~ {\rm or} ~ P_b \equiv 0 ,&  J +P_b 
\equiv (m+ n) L  +Q , ~  m L + r \pmod{N}. 
\elea(Pab)
We define the $t^N$-polynomial (\cite{R0805} (4.24) (4.35)):
\bea(ll)
P(t) &= C t^{-P_a-P_b} \frac{\tau^{(N)}(t)}{F(t)^2} \\
&= C \omega^{-P_b} \sum_{k=0}^{N-1} \frac{h(t) \cdots h( \omega^{k-1}t) h( \omega^{k+1}t)\cdots h( \omega^{N-1}t) (\omega^k t)^{-(P_a+P_b)}}{F(\omega^k t) F( \omega^{k+1} t) }
\elea(Pdef)
with $P ( 0 ) \neq 0$, where $C = \omega^\frac{-(1+2m+n)(N-1)L}{2} \prod_{\ell=1}^L t_{p_\ell}^{N-1}$. Using the coordinates $(t, \lambda)$ in (\req(hW)), one can factorize $P(t)$ using a $\lambda$-function $G$:
\be
P (t) = D G (\lambda) G (\lambda^\dagger),  
\ele(Pt)
where $D$ is the $q$-independent function in (\req(TTD)). Then the  eigenvalues  of the normalized transfer matrices $V, \widehat{V}$ of $T, \widehat{T}$ with $\widehat{V} = V D$ ( \cite{B90} Sect. 2, \cite{R0805} (4.4) ):
$$
V (q) =  T (q) \bigg(  \mu_q^\frac{N(N-1)L}{2} \prod_{\ell=1}^L g_{p_\ell} (q) \overline{g}_{p'_\ell} (q) \bigg)^{\frac{-1}{N}} , ~ ~
\widehat{V} (q) =  \widehat{T} (q) \bigg(  \mu_q^\frac{N(N-1)L}{2} \prod_{\ell=1}^L g_{p'_\ell} (q) \overline{g}_{p_\ell} (q) \bigg)^{\frac{-1}{N}} ,
$$
are expressed by (\cite{R0805} (4.34))
\be
V (q) =   \zeta_0^\frac{L}{N} x_q^{P_a } y_q^{P_b }  \mu_q^{-P_\mu}   
 \frac{F( t_q )}{\prod_{\ell =1}^L \prod_{k=1}^{N-1} (t_{p'_\ell} - \omega^k t_q)^\frac{k}{N}}  G( \lambda_q ), 
\ele(Vq)
where $\zeta_0 = e^\frac{ \pi {\rm i}(N-1)(N+4)}{12}$, and $P_\mu \equiv r \pmod{N}$. In particular when $r=0$ and $m=n=0$ in (\req(supdef)), an equivalent expression of $T$-eigenvalue appeared in \cite{ND08} section 3.2 through the Algebraic-Bethe-Ansatz method.

\subsection{Onsager-algebra symmetry and the induced $sl_2$-loop-algebra structure in homogeneous superintegrable chiral Potts model \label{ssec.OACP}}
We now consider the homogeneous superintegrable case, i.e. $p= p_\ell = p_\ell^\prime$ for all $\ell$ in (\req(supdef)). The $L$-operators (\req(tau2)) are all equivalent to 
\be
L_\ell ( t ) = \left( \begin{array}{cc}
        1  -  \omega^n  {\tt t} X   &  (1  -\omega^{1+m+n}  X) Z \\
       - {\tt t} ( 1  -  \omega^{m+n}  X )Z^{-1} & - {\tt t} +  \omega^{1+2m+n}  X
\end{array} \right) ~ ~ ~ ( {\tt t} := \omega^m t_p^{-1} t ) ~ ~ ~ {\rm for \ all \ } \ell . 
\ele(hsupL)
Using the above normalized spectral parameter ${\tt t}$, the $\tau^{(2)}$-models for all $k'$ are the same when the vertical rapidities $p, p'$ lie in a curve ${\goth W}$ in (\req(cpmC)). Express all polynomials in (\req(tausp)) in terms of ${\tt t}$:
\bea(ll)
h(t) = {\tt h} ({\tt t}) = (1- \omega^{-m} {\tt t})^L , & F (t) = {\tt F}( {\tt t} ) = \prod_{j=1}^J (1+ \omega {\tt v}_j {\tt t}) ~ ~ ~ ({\tt v}_j := \omega^{-m} t_p v_j ) .
\elea(hFv)
The Bethe equation (\req(Besupin)) becomes (\cite{R0805} (4.31) (4.32)):
\be
 ( \frac{{\tt v}_i  + \omega^{-1-m}  }{{\tt v}_i  + \omega^{-2-m} })^L   = - \omega^{-P_a- P_b} \prod_{j=1}^J \frac{{\tt v}_i -  \omega^{-1}   {\tt v}_j }{ {\tt v}_i -\omega  {\tt v}_j } , \ \ i= 1, \ldots, J,
\ele(Bethesup)
with $\tau^{(2)}$-eigenvalues (\req(tausp)) expressed by
\be
\tau^{(2)}(t) = \omega^{-P_a} (1- \omega^{-m} {\tt t})^L \frac{ {\tt F} ({\tt t})}{{\tt F}( \omega {\tt t})} + \omega^{P_b}(1- \omega^{1-m} {\tt t})^L \frac{{\tt F} (\omega^2 {\tt t})}{{\tt F} ( \omega {\tt t})}  .
\ele(stauev)
Normalize $P(t)$ in (\req(Pdef)) by 
\bea(ll)
P(t) = C_p {\tt P}( {\tt t} ), &{\tt P}({\tt t}) = \omega^{-P_b}  \sum_{k=0}^{N-1} \frac{(1-  {\tt t}^N )^L (\omega^{k} {\tt t})^{-(P_a+P_b)}}{(1- \omega^{-m+k} {\tt t})^L {\tt F} (\omega^k {\tt t}) {\tt F}( \omega^{k+1} {\tt t}) } .
\elea(sPt)
where $C_p = \omega^{\frac{-(1+2m+n)(N-1)L}{2}+m (P_a+P_b)}  t_p^{L(N-1)-P_a-P_b}$. Note that the Bethe relation (\req(Bethesup)) is the polynomial criterion of ${\tt P}( {\tt t} )$ in (\req(sPt)), which can be regarded as a ${\tt t}^N$-polynomial with the degree 
$$
m_E:= [\frac{(N-1)L-P_a-P_b-2J}{N}]
$$ 
with ${\tt P}(0) \neq 0$. We shall denote the $\tau^{(2)}$-eigenspace with the eigenvalue (\req(stauev)) for a Bethe polynomial ${\tt F}({\tt t)}$ in (\req(Bethesup)) by ${\cal E}_{{\tt F}, P_a, P_b}$. Then the dimension of ${\cal E}_{{\tt F}, P_a, P_b}$ is equal to $2^{m_E}$. One has the $\tau^{(2)}$-eigenspace-decomposition of the quantum space:
\be
\bigotimes^L \CZ^N = \bigoplus\{ {\cal E}_{{\tt F}, P_a, P_b}| ~ {\tt F}: {\rm Bethe ~ polynomial ~with ~ quantum ~ numbers} ~ P_a, P_b \} .
\ele(EF)

For simple notations, hereafter  we assume the vertical rapidity $p$ of a homogeneous superintegrable CPM always in  ${\goth W}_{k'} ~(k' \neq 0 , \pm 1) $ with $\mu_p ~(=\omega^\frac{n}{2})$ being a $N$th root of unity, equivalently $n \equiv 2n_0$ for some $n_0 \in \ZZ$, (a constraint required only in even $N$ case)\footnote{In the even-$N$ and odd-$n$ case, $\mu_p^N= -1$ with the $\eta$ in (\req(pcood)) changed to $\eta^{-1}$ in the definition of $p$. One may also discuss the Onsager-algebra symmetry of superintegrable CPM by changing $\eta$ here to $\eta^{-1}$ in the argument.}. First we consider the case 
\bea(ll)
p: (x_p, y_p, \mu_p)= (\eta^{\frac{1}{2}}\omega^m,  \eta^{\frac{1}{2}}, \omega^{n_0}) \in {\goth W}_{k'}, & \eta := (\frac{1-k'}{1+k'})^{\frac{1}{N}}, 
\elea(pcood)
where $k' \neq 0 , \pm 1$, and $0\leq m \leq N-1$ (see \cite{AMP, B93, B94, R05o, R075}). All $\tau^{(2)}, T , \widehat{T}$-matrices commute with each other for  $q \in {\goth W}_{k'}$. Using (\req(gg1)) (\req(Pt)) (\req(Vq)) and (\req(sPt)), one finds the following formulas of $T, \widehat{T}$-eigenvalues:
\bea(ll)
T(q) = \alpha_1 N^L \frac{R_m ( {\tt x} )^L ( 1-  {\tt x} )^L}{R_m ({\tt y})^L( 1-   {\tt x}^N )^L}  {\tt x}^{P_a }{\tt y}^{P_b }  \mu^{-P_\mu} \frac{{\tt F}({\tt t} )}{\omega^{P_b+m(P_b +P_a)}{\tt F}( \omega^{m+1}) } {\cal G}(\lambda) ,  \\
\widehat{T} (q) 
= \alpha_1^{-1} N^L \frac{R_m ( {\tt x})^L ( 1- {\tt x} )^L}{R_m ({\tt y})^L( 1- {\tt x}^N )^L}  {\tt x}^{P_a } {\tt y}^{P_b }  \mu^{-P_\mu} \frac{{\tt F}({\tt t} )}{{\tt F}(\omega^m) } {\cal G}(\lambda),  
\elea(TTform)
where $\alpha_1 := (-1)^{mL} \omega^{\frac{m(m+1) L+ 2m P_a - 2n_0 P_\mu  }{2}}$, $R_m (z):= \frac{( 1- z^N )}{\prod_{j=0}^{N-1-m} (1- \omega^j z )}$, $\mu:= \mu_q, ~ \lambda:= \mu^N$, the variables ${\tt x}, {\tt y}, {\tt t}$ are the normalized coordinates of $x_q , y_q, t_q$:
\be
{\tt x} := \omega^m x_p^{-1} x_q , ~ ~ {\tt y} := y_p^{-1} y_q , ~ ~ {\tt t} := \omega^m t_p^{-1} t_q , 
\ele(xyt)
and ${\cal G}(\lambda)$ is the $\lambda$-function on $W_{k'}$ in (\req(hW)) to factorize the polynomial ${\tt P}({\tt t})$ in (\req(sPt)):  ${\cal G} (\lambda) {\cal G} (\lambda^{-1}) = \frac{{\tt P}({\tt t})}{{\tt P}(\omega^m )}$.
Note that $\widehat{T} (p) = 1$, and the total momentum $S_R$ (=$D$ in (\req(Pt))) is defined by
\be
 S_R = \omega^{- m(m+1) L +m(P_b -P_a)+ 2n_0 P_\mu} \frac{\omega^{P_b} F( \omega t_p)}{F(t_p)}.
\ele(SR)
The function ${\cal G}(\lambda)$ is related to $G(\lambda_q)$ in (\req(Vq)) by 
\be
G(\lambda_q)  = \alpha_1  e^{\frac{-\pi {\rm i}(1+2n_0)(N-1)L}{2N}}   N^{\frac{L}{2}} (\frac{1- {\tt x}^N}{1- {\tt y}^N})^\frac{mL}{N} \frac{\eta^{\frac{1}{2}(L(N-1)-P_a-P_b)}}{\omega^{P_b+m(P_b +P_a)} F(\omega^{1+m} \eta) } {\cal G}(\lambda), 
\ele(GG)
where $\lambda = \lambda_q$. 
Since the Boltzmann weights are finite when any of $x_q, y_q, \mu_q, \mu_q^{-1}$ tends to zero, it follows $P_a, P_b$ being non-negative integers, and  $P_b + J - mL \leq P_\mu \leq (N-1-m)L -P_a - Nm_E-J$. 
In the case when $m=0$, the formula (\req(TTform)) is the same as that in \cite{AMP, B89, B93, B94}, and the relation (\req(GG)) has been given in \cite{R0805} section 4.3\footnote{For the ${\tt m}=k={\tt n}=0$ case in \cite{R0805} section 4.3, some misprint occurred in the formulas of $T(q), V(x_q, y_q),S(\lambda_q)$ there, where $F( \eta^{-1} t_q), F(\omega^k \eta^{-1} t_q)$  should be $F( \eta^{-1} t_q), F(\omega^k \eta^{-1} t_q)$ respectively.}. Therefore ${\cal G} (\lambda)$ are determined by the zeros of ${\tt t}^N$-polynomial ${\tt P}({\tt t})$, denoted by ${\tt t}^N_i, ~ i=1, \ldots, m_E,$ through $W_{k'}$ in (\req(hW)), equivalently the curve 
\bea(lll)
W_{k'}&:  \frac{(1-k')^2 }{4} w^2 = \frac{(1-k')^2 }{4} + \frac{k'}{1- {\tt t}^N }, &   (w:= \frac{\lambda + 1}{\lambda- 1}, ~ {\tt t} := \omega^m t_p^{-1} t) .
\elea(hWw)
Indeed, ${\cal G}(\lambda)$ is expressed by
\be
{\cal G}(\lambda)= \prod_{i=1}^{m_E} \frac{(\lambda +1) - (\lambda -1) w_i}{2 \lambda }
\ele(cG)
where $w_i$'s are solutions in equation (\req(hWw)) for ${\tt t}^N= {\tt t}^N_i$ (\cite{AMP} (2.22), \cite{B94} (20)). 
There are two solution of $w_i = \pm \overline{w}_i $, where $\Re (\frac{1-k'}{2} \overline{w}_i) > 0$ for real $k'> 0$. 
 Any choice of $w_i=  s_i \overline{w}_i ~ (1 \leq i \leq m_E)$ with $s_i = \pm$ gives rise to a $T$-(or $\widehat{T}$)-eigenvalue (\req(TTform)) with the norm-one eigenvector, denoted by $\vec{v}(s_1, \ldots, s_{m_E} ; k')$. All such vectors form a basis of the $\tau^{(2)}$-eigenspace ${\cal E}_{{\tt F}, P_a, P_b}$ in (\req(EF)):
\be
{\cal E}_{{\tt F}, P_a, P_b}= \bigoplus_{s_i = \pm} \CZ ~ \vec{v}(s_1, \ldots, s_{m_E} ; k') .
\ele(Ek')
Note that when $s_E=0$, ${\cal E}_{{\tt F}, P_a, P_b}$ is of dimension 1 with $\vec{v}(s_1, \ldots, s_{m_E} ; k')$ being the norm-one base element. Furthermore, the integers $P_a, P_b$ and $P_\mu$ in (\req(TTform)) are indeed quantum numbers of $\tau^{(2)}$-model, depending only on the $\tau^{(2)}$-eigenvalue. 

We now discuss the Onsager algebra $({\sf OA})$ and $sl_2$-loop algebra $(sl_2[z, z^{-1}])$ symmetry of a homogeneous superintegrable CPM with the boundary condition $r$.
As $q$ tends to $p$ in ${\goth W}_{k'}$, up to the first order with small $\epsilon$, we may set 
\bea(lll)
x_q = \omega^m \eta^{\frac{1}{2}} (1- 2k' \epsilon ), & y_q = \eta^{\frac{1}{2}}(1 + 2k' \epsilon ) , & \mu_q = \omega^\frac{n}{2}( 1+ 2(k'-1) \epsilon). 
\elea(xyep)
Then $\widehat{T} (q)$ near $p$ ( \cite{AMP} (1.11)-(1.17) )\footnote{The variable $\epsilon$ is related to the $u$ in \cite{AMP} (1.11) (or (\req(uapr)) in this paper) by  $(-1)^m u = \epsilon$. Note that in the even $N$ case when $\mu_p^N= -1$, the expression of $\widehat{T} (q)$ near $p$ here still holds by changing $H (k')$ to $H (-k')$.} is expressed by 
$$
\widehat{T} (q) = {\bf 1} \{ 1 +   (N-1-2m) L \epsilon \} + \epsilon H (k') + O(\epsilon^2) 
$$
where the Hamiltonian $H (k') = H_0 + k' H_1 $ is expressed by
\bea(ll)
H_0 = - 2\sum_{\ell =1}^L  \sum_{j=1}^{N-1}   \frac{ \omega^{m j }Z^j_\ell Z^{-j}_{\ell +1} }{1-\omega^{-j}}& 
H_1 = - 2 \sum_{\ell =1}^L \sum_{j=1}^{N-1}    \frac{ \omega^{(m + 2n_0 )j }X_\ell^j }{1-\omega^{-j}}.
\elea(H01)
with the boundary condition: $Z_{L+1}= \omega^{-r} Z_1, ~ X_{L+1}= X_1$. By (\req(TTform)), one may regard (\req(Ek')) as the $H (k')$-eigenvector decomposition of ${\cal E}_{{\tt F}, P_a, P_b}$ with the $H (k')$-eigenvalues:
\bea(l)
E(s_1, \ldots, s_{m_E}; k') =  2 P_\mu + Nm_E - (N-1-2m) L  \\
+k'\bigg( (N-1- 2m)L -N m_E + 2(P_b - P_a - P_\mu)  \bigg) + 2N \sum_{i=1}^{m_E} \frac{(1-k')}{2}(s_i \overline{w}_i).
\elea(Esk')
(When $s_E=0$, $\sum_{i=1}^{m_E} $ in (\req(Esk')) and $\prod_{i=1}^{m_E} $ in (\req(cG)) are defined to be $0, 1$ respectively.)
The matrices $\tau^{(2)} (t)$ are the same for all $k'$ when using the rescaled variable ${\tt t}$ in (\req(hsupL)), hence commute with $H (k')$ for all $k'$, equivalently, $[\tau^{(2)} (t), H_0] = [\tau^{(2)} (t), H_1]=0$. Since the operators $H_0, H_1$ in (\req(H01)) satisfy the Dolan-Grady relation for the Onsager-algebra generators \cite{Dav, GR, R075}, one obtains an ${\sf OA}$ representation on the $\tau^{(2)}$-eigenspace ${\cal E}_{{\tt F}, P_a, P_b}$ in ({\req(EF)) \cite{R05o, R075}. By the general theory of ${\sf OA}$ representation, this Onsager-algebra symmetry is inherited from a $sl_2$-loop-algebra structure of ${\cal E}_{{\tt F}, P_a, P_b}$  \cite{R91}. 
We now describe their explicit relationship. As it was known in \cite{R91}, ${\sf OA}$ can be regarded as the Lie subalgebra of $sl_2[z, z^{-1}]$ fixed by the involution, $\iota: (e^+, h, z) \leftrightarrow (e^-, -h, z^{-1})$, where $e^\pm, h$ are the standard generators of $sl_2$ with $[e^+, e^-]=H, [h, e^\pm]= \pm 2e^\pm$; and a finite-dimensional representation of ${\sf OA}$ can be factored through a $sl_2[z, z^{-1}]$-representation. As before we denote by ${\tt t}^N_i ~ (i=1, \ldots, m_E)$  the zeros  of ${\tt t}^N$-polynomial ${\tt P}({\tt t})$, and write $\cos \theta_i = \frac{1+ {\tt t}^{-N}_i}{1-{\tt t}^{-N}_i}$, equivalently 
\bea(lll)
{\rm e}^{{\rm i} \theta_i}=   \frac{1+ {\tt t}_i^{-N/2} }{1- {\tt t}_i^{-N/2}}  & (\leftrightarrow {\tt t}^{-N}_i =(\frac{{\rm e}^{{\rm i} \theta_i}-1}{{\rm e}^{{\rm i} \theta_i}+1})^2) , & \Im ({\tt t}_i^{-N/2}) \geq 0.
\elea(OAev)
By \cite{R91} (Theorem 3 and (32) (33)), there exists a basis of ${\cal E}_{{\tt F}, P_a, P_b}$, $\vec{\bf b}(s_1,  \ldots, s_{m_E}) ~ $ for $s_i = \pm $ such that the generator $H_0, H_1$ of the ${\sf OA}$-representation ${\cal E}_{{\tt F}, P_a, P_b}$  can be expressed in the form: 
\bea(lll)
H_0 &= \alpha + N \sum_{i=1}^{m_E} (e^+_i + e^-_i )&= \alpha +  2 N \sum_{i=1}^{m_E} J^x_i , \\
H_1 &=  \beta - N \sum_{i=1}^{m_E} ({\rm e}^{{\rm i} \theta_i} e^+_i  +{\rm e}^{-{\rm i} \theta_i} e^-_i  )&=  \beta -  2 N \sum_{i=1}^{m_E} ( \cos \theta_i J^x_i - \sin \theta_i J^y_i), 
\elea(H01eh)
where $e^\pm_i, h_i$ are the $sl_2$-generators for the basis elements $\vec{\bf b}(s_i, \ldots, s_i, \ldots, s_{m_E})$ acted only on the $i$th $s_i = \pm $ as the spin-$\frac{1}{2}$ representation, and  $
J^x_i := \frac{1}{2}(e^+_i+e^-_i), J^y_i := \frac{-{\rm i}}{2}(e^+_i-e^-_i), J^z_i := \frac{1}{2}h_i$. The ${\sf OA}$-representation (\req(H01eh)) is induced from the $sl_2[z, z^{-1}]$-structure of ${\cal E}_{{\tt F}, P_a, P_b}$ by the evaluating $z$ at ${\rm e}^{{\rm i} \theta_i}$'s:
\be
\varrho: sl_2[z, z^{-1}] \longrightarrow {\rm GL} ({\cal E}_{{\tt F}, P_a, P_b}), ~ ~ ~ e^\pm z^k \mapsto \sum_{i=1}^{m_E} e_i^\pm {\rm e}^{{\rm i} \theta_i k} .
\ele(rho)
By (\req(Esk')) and (\req(H01eh)),  $
H(k') = \alpha + k' \beta + 2N \sum_{i=1}^{m_E} \bigg( (1- k' \cos \theta_i) J^x_i + k' \sin \theta_i J^y_i \bigg)$, 
with  eigenvalues $\alpha + k' \beta + N \sum_{i=1}^{m_E} \pm \sqrt{1+k'^2 - 2k' \cos \theta_i}$, where 
$\alpha = 2 P_\mu + Nm_E - (N-1-2m) L$, $\beta = (N-1- 2m)L -N m_E + 2(P_b - P_a - P_\mu)$, and $
\sqrt{1+k'^2 - 2k' \cos \theta_i} = (1-k')\overline{w}_i$. 
Furthermore, the $H(k')$-eigenvectors in (\req(Ek')) are related to $\vec{\bf b}(s_1,  \ldots, s_{m_E})$ by
\be
\vec{v}(s_1, \ldots, s_{m_E} ; k') = (2{\rm i})^{-m_E/2} (\prod_{i} s_i )   \sum_{s_1', \ldots, s_{m_E}'} ( \prod_i  (\sqrt{s_i} {\rm e}^{{\rm i}  \varphi_i/2})^{s_i'} ) 
\vec{\bf b}(s_1', \ldots, s_{m_E}')
\ele(vb)
where ${\rm e}^{{\rm i} \varphi_i} = \frac{1- k' \cos \theta_i - {\rm i} k' \sin \theta_i}{\sqrt{1+k'^2 - 2k' \cos \theta_i}}$, and we make the identification $\pm = \pm 1$, $\sqrt{+} := 1, \sqrt{-} := {\rm i}$. Note that for a given $k'$, there is a $(\stackrel{m_E}{\oplus} sl_2)$-algebra structure of ${\cal E}_{{\tt F}, P_a, P_b}$ with the generators $e^\pm_{i, k'}, h_{i,k'}$, (or $J^x_{i, k'}, J^y_{i, k'} J^z_{i, k'}$) acting on the basis elements $\vec{v}(s_1, \ldots, s_{m_E} ; k')$ only on the $i$th $s_i$. One may also introduce the $k'$th $sl_2[z, z^{-1}]$-structures of ${\cal E}_{{\tt F}, P_a, P_b}$ by
\be
\rho_{k'} : sl_2[z, z^{-1}] \longrightarrow {\rm GL} ({\cal E}_{{\tt F}, P_a, P_b}), ~ ~ ~ e^\pm z^k \mapsto \sum_{i=1}^{m_E} e_{i, k'}^\pm {\rm e}^{{\rm i} \theta_i k} .
\ele(rhok') 
By (\req(vb)), the following operators of ${\cal E}_{{\tt F}, P_a, P_b}$ are identical:
$$
\begin{array}{lll}
J^x_i = \cos \varphi_i J^z_{i, k'} + \sin \varphi_i J^x_{i, k'}, & J^y_i  = - \sin \varphi_i J^z_{i, k'} + \cos \varphi_i J^x_{i, k'}, & J^z_i = J^y_{i, k'} 
\end{array}
$$
equivalently,
$$ 
\begin{array}{lll}
\cos \varphi_i J^x_i - \sin \varphi_i J^y_i  = J^z_{i, k'}, & \sin \varphi_i J^x_i + \cos \varphi_i J^y_i  = J^x_{i, k'}, & J^z_i = J^y_{i, k'} ,
\end{array}
$$
which in turn yield $H(k') = \alpha + k' \beta + 2N \sum_{i=1}^{m_E}  \sqrt{1+k'^2 - 2k' \cos \theta_i} J^z_{i, k'}$. In particular as $k'$ tends $0$ or $\infty$, one obtains
$$
\begin{array}{lll}
J^x_i =  J^z_{i, 0} ,&  J^y_i  =  J^x_{i, 0}, & J^z_i = J^y_{i, 0} \\
- \cos \theta_i J^x_i + \sin \theta_i J^y_i  = J^z_{i, \infty},&  - \sin \theta_i J^x_i - \cos \theta_i J^y_i  = J^x_{i, \infty},&  J^z_i = J^y_{i, \infty} . 
\end{array}
$$
Using the above relations, one finds the equivalent $sl_2[z, z^{-1}]$-structure on ${\cal E}_{{\tt F}, P_a, P_b}$ induced from $\rho_{k'}~ (k'=0, \infty)$ and the standard $\varrho$ in (\req(rho)):
$$
\varrho = \rho_{0}  \cdot R , ~ ~ \varrho  = \rho_{\infty}\cdot R \cdot \iota \cdot  \nu ,
$$   
where $R$ and (involutions) $\iota, \nu$  are $sl_2[z, z^{-1}]$-automorphisms defined by 
$$
\begin{array}{l}
R:(J^x, J^y, J^z , z) \mapsto (J^z, J^x, J^y, z), ~ ~ \iota: (e^\pm , h , z ) \mapsto (e^\mp, -h, - z^{-1}), \\
\nu : (e^+, e^-, z^{-1} e^-, z e^+,  h ) \mapsto (ze^-, z^{-1}e^+, e^+, e^-, -h).
\end{array}
$$
Later in section \ref{ssec.symEq}, we shall compare the above $sl_2[z, z^{-1}]$-structure of ${\cal E}_{{\tt F}, P_a, P_b}$ with the $sl_2$-loop-algebra symmetry induced from the XXZ chain equivalent to the $\tau^{(2)}$-model in \cite{R0806}. \par \vspace{1mm} \noindent
{\bf Remark.} (1) The discussion of this subsection is still valid for the homogeneous superintegrable CPM with a vertical rapidity $p= (\eta^\frac{1}{2}\omega^{m+m'}, \eta^\frac{1}{2}\omega^{m'}, \omega^{n_0})$. Indeed, under the automorphism $(x, y, \mu) \mapsto (\omega^{m'}x, \omega^{m'}y, \mu)$ on both vertical and horizontal rapidities, the Boltzmann weights (\req(Weig)) and the chiral Potts transfer matrix (\req(ThatT)) are unchanged, so are the values of ${\tt x, y, t}-$coordinates in (\req(xyt)); hence the results and formulas of this subsection remain the same. \par \noindent
(2) The ${\sf OA}$-representation of ${\cal E}_{{\tt F}, P_a, P_b}$ in (\req(H01eh)) is indeed irreducible. By the theory of ${\sf OA}$-representation (\cite{R05o} \cite{DR} Theorem 6), the evaluated values $e^{{\rm i} \theta_i}$'s in (\req(OAev)) for the $sl_2[z, z^{-1}]$-representation (\req(rho)) satisfy the relation, $
e^{{\rm i} \theta_i} \neq 0, \pm 1$ and $e^{{\rm i} \theta_i} \neq e^{\pm {\rm i} \theta_j}~ (i \neq j)$, equivalently    ${\tt t}_i^N \neq 1, 0$, ${\tt t}_i^N \neq  {\tt t}_j^N ~ (i \neq j)$ for the roots of ${\tt P}({\tt t})$.

\section{Duality in Chiral Potts Model and $\tau^{(2)}$-model \label{sec.DualCP}}
\setcounter{equation}{0}
In this section, we derive the duality in  $\tau^{(2)}$-model and chiral Potts model. For simpler notations, we discuss only the alternating-rapidity case, i.e. the rapidities in (\req(pp'l)) at  all sites  are the same:
\be
p= p_\ell, ~ ~ p'= p'_\ell ~ \in {\goth W} ~ ~ {\rm for ~ all } ~ ~ \ell 
\ele(pp')
where ${\goth W}$ is a curve in (\req(cpmC)). Note that with the same argument, all formulas in this section still hold for the inhomogeneous case after a suitable modification of scalar coefficients, as illustrated in section \ref{sec:CPM} in comparison with the parallel results in \cite{R0805}.

\subsection{Duality in $\tau^{(2)}$-model and $\tau^{(2)}$-face-model  \label{ssec.Ftau}}
In the alternating-rapidity case (\req(pp')), the parameter of $L_\ell$ in (\req(L)) is given by
\bea(l)
({\tt a}, {\tt a}', {\tt b}, {\tt b}', {\tt c}) = (x_p, x_{p'}, y_p, y_{p'}, \mu_p \mu_{p'}),  
\elea(Lpp')
and we write $\tau^{(2)}$-matrix in (\req(t2inh)) simply by $\tau^{(2)}(t) = \tau^{(2)}(t; p,p')$. It is known that $\tau^{(2)}(t)$ can be written as the following product form (\cite{BBP} $(3.44a)_{k=0, j=2}$ (3.48),  \cite{B93} (2.14)-(2.16)):
\be
\tau^{(2)}(t; p,p')_{\{\sigma \}, \{\sigma''\}} =  \prod_{\ell =1}^L U(\sigma_\ell,\sigma''_\ell | \sigma_{\ell+1},\sigma''_{\ell+1} ).  
\ele(taupd)
Here the factor $U(a, d| b, c)$ is defined by
\be
U(a, d| b, c) = \sum_{m=0, 1} \omega^{m(d-b)} (-\omega t)^{a-d-m} F_p(a-d , m) F_{p'}(b-c, m) 
\ele(Uf)
where $F_p(0, 0)=1$, $F_p(0, 1)= \frac{- \omega t}{y_p}$, $F_p(1, 0)=  \frac{\mu_p }{y_p}$, $F_p(1, 1)=  \frac{ -\omega x_p \mu_p }{y_p}$, and $F_p(\alpha, m)= 0$ if $\alpha \neq  0 , 1$. One may express (\req(Uf)) in terms of the "face-variables", $n:= a-b, ~ n':= d-c \in \ZZ_N$, and form the ${\tt L}$-operator of $\tau^{(2)}$-face model:
\bea(l)
{\tt L}_\ell ( t ) = \left( \begin{array}{cc}
        1  -  t \frac{1 }{y_p y_{p'}} {\tt Z}   & (\frac{\mu_{p'}}{y_{p'}}  -\omega   \frac{x_{p'}\mu_{p'} }{y_py_{p'}} {\tt Z}) {\tt X}^{-1} \\
       - t  (\frac{\mu_p}{y_p} -  \frac{x_p\mu_p}{y_p y_{p'}}{\tt Z}){\tt X} & - t \frac{\mu_p \mu_{p'}}{ y_p y_{p'}}+  \omega \frac{x_p x_{p'} \mu_p \mu_{p'}}{ y_p y_{p'}}{\tt Z} 
\end{array} \right), ~ 1 \leq \ell \leq L ,
\elea(FaLpp')
where ${\tt X}, {\tt Z}$ are the Weyl-operators of the face-quantum space ${\tt C}^N := \sum_{n \in \ZZ_N} \CZ |n \rangle \rangle $,  
 ${\tt X}|n' \rangle \rangle= |n'+1 \rangle \rangle, {\tt Z}|n' \rangle \rangle = \omega^{n'} |n' \rangle \rangle$. Under the identification of $({\tt X}, {\tt Z})$ with $(\widehat{X}, \widehat{Z})$, the ${\tt L}$-operator (\req(FaLpp')) is the same as the $L$-operator (\req(L)) with the parameter
\bea(l)
({\tt a}^*, {\tt a}'^*, {\tt b}^*, {\tt b}'^*, {\tt c}^*) = (x_{p'}\mu_{p'}, x_p\mu_p, \frac{y_{p'}}{\mu_{p'}}, \frac{y_p}{\mu_p}, \frac{1}{\mu_{p'}\mu_p }).  
\elea(fLpp')
Therefore  (\req(FaLpp')) satisfies the YB relation (\req(YBtau)), so is the monodromy matrix
\be
{\tt L}_1 (t) {\tt L}_2 (t) \cdots {\tt L}_L (t) = \left( \begin{array}{cc}
        {\tt A}(t)  & {\tt B}(t) \\
        {\tt C}(t) &  {\tt D}(t)
\end{array} \right).  
\ele(MtF)
The face $\tau^{(2)}$-operator with the boundary condition
$$
| n_{L+1} \rangle \rangle  = \omega^{-{\tt r}n_1} | n_1 \rangle \rangle , ~ ~ ({\tt r} \in \ZZ_N )
$$ 
is the commuting family defined by 
\be
\tau_F^{(2)}(t) (= \tau_F^{(2)}(t; p, p')) = {\tt A}(\omega t) + \omega^r {\tt D}(\omega  t) . 
\ele(Ftau2) 
It is easy to see that $\tau_F^{(2)}(t)$ commute with the charge operator ${\tt Z} ~ (:= \prod_{\ell} {\tt Z}_\ell)$, whose eigenvalues are $\omega^{\tt Q}$ for ${\tt Q} \in \ZZ_N$.

Define the birational morphism of $\CZ^3$,
\be
p= (x_p, y_p, \mu_p) \longrightarrow  p^* = (x_{p*}, y_{p*}, \mu_{p*}):= ( {\rm i}^\frac{1}{N} x_p \mu_p, {\rm i}^\frac{1}{N} y_p \mu_p^{-1}, \mu_p^{-1}), 
\ele(pp*)
which identifies the chiral Potts curves of $k'$ and $k'^{-1}$, ${\goth W} \cong {\goth W}^*$, with ${\goth W}$ in (\req(cpmC)) and its dual curve defined by
\bea(l)
{\goth W}^* = \left\{\begin{array}{ll}
{\goth W}_{\frac{1}{k'}, \frac{{\rm i}k}{k'}} = {\goth W}_\frac{1}{k'} & {\rm  if} ~ {\goth W}= {\goth W}_{k', k} = {\goth W}_{k}, ~ ( k'^2 \neq 1, 0), \\
{\goth W}_{\pm 1}^*,  \overline{\goth W}_{\pm 1}^*,  & {\rm  if} ~ {\goth W}= {\goth W}_{\pm 1}, \overline{\goth W}_{\pm 1} ~ {\rm respectively} .
\end{array}
\right.
\elea(W*)   
where ${\goth W}_{\pm 1}^*,  \overline{\goth W}_{\pm 1}^*$ are obtained by the substitution of variables, $(x, y, \mu) = ({\rm i}^\frac{1}{N} x^*, {\rm i}^\frac{1}{N} y^*, \mu^*)$, in the equation of ${\goth W}_{\pm 1}, \overline{\goth W}_{\pm 1}$ respectively. Note that for $k' \neq \pm 1$, $p \mapsto p^{* *}$ is the canonical identification between ${\goth W}_{k', k}$ and ${\goth W}_{k', -k}$. For $p \in {\goth W}$, the corresponding $p^*$ in the dual curve ${\goth W}^*$ will be called the dual rapidity of $p$. By (\req(FaLpp')) and (\req(fLpp')), the ${\tt L}$-operator of $\tau_F^{(2)}(t; p, p')$ is gauge-equivalent to  the $L$-operator of $\tau^{(2)}(t^*; p'^*, p^*)$ via the diagonal matrix ${\rm dia}[1, {\rm i}^\frac{1}{N}]$. Hence the face $\tau^{(2)}$-model $\tau_F^{(2)}(t; p, p')$ is equivalent to $\tau^{(2)}(t^*; p'^*, p^*)$ with the boundary condition $r^*$ via the identification  of quantum vector spaces, 
\be
\Phi : {\tt C}^N \rightarrow \CZ^N, ~ ~ |n \rangle \rangle \mapsto \widehat{|n} \rangle,
\ele(Phi) 
and the identical boundary and charges condition, $({\tt r}, {\tt Q}) = (r^*, Q^*)$:
\be 
 \tau_F^{(2)}(t; p, p') = \Phi^{-1} \tau^{(2)}(t^*; p'^*, p^*)  \Phi , ~ ~ t^*= (-1)^\frac{1}{N} t .
\ele(tf=t)
On the other hand, the $\tau^{(2)}$-models, (\req(taupd)) and (\req(Ftau2)), are closely related through the correspondence of their quantum spaces:
$$
\begin{array}{ll}
\Theta_r (= \Theta ): \bigotimes^L \CZ^N \longrightarrow \bigotimes^L {\tt C}^N, & |\sigma_1, \ldots \sigma_L \rangle \mapsto |n_1, \ldots n_L \rangle \rangle ,  n_\ell := \sigma_\ell - \sigma_{\ell +1} ,
\end{array}
$$
where $1 \leq \ell \leq L$ with $\sigma_{L +1} \equiv \sigma_1 - r$. 
It is easy to see that the kernel of $\Theta_r = \CZ^N$ with the image being the ${\tt Z}$-charge-$r$ subspace. Indeed, the inverse of the vector $|n_1, \ldots n_L \rangle \rangle$ is expressed by
$$
\Theta_r^{-1} ( |n_1, \ldots n_L \rangle \rangle )= \bigoplus_{Q=0}^{N-1} |Q; n_1, \ldots n_L \rangle
$$
where $| Q; n_1, \ldots n_L \rangle =  N^{-1/2} \sum_{\sigma_1=0}^{N-1} \omega^{-Q \sigma_1} |\sigma_1, \ldots \sigma_L \rangle$ with $\sigma_\ell - \sigma_{\ell +1} = n_\ell$ \cite{AuP7}. For a fixed $Q$, all $|Q; n_1, \ldots n_L \rangle$ with $\sum_{\ell=1}^L n_\ell = r$ form a basis of the subspace 
\be
V_{r, Q} = \{ v= ( v_{\sigma_1, \ldots \sigma_L}) \in \bigotimes^L \CZ^N| ~ \sigma_{L+1} \equiv \sigma_1 - r , ~ ~ X (v) = \omega^Q v \} ,
\ele(VrQ)
with the dual basis, $\langle Q; n_1, \ldots n_L | =  N^{-1/2} \sum_{\sigma_1=0}^{N-1}\omega^{Q \sigma_1} \langle \sigma_1, \ldots \sigma_L  |$,  $(\sigma_\ell - \sigma_{\ell +1} = n_\ell)$.
Under $\Theta_r$, the vector space $V_{r, Q}$ is isomorphic to the $\Theta_r$-image, regarded as a subspace of the quantum space  $\bigotimes^L {\tt C}^N$ with ${\tt r}\equiv Q$, and ${\tt Q} \equiv r$:
$$
W_{{\tt r}, {\tt Q}} = \{ w= ( w_{n_1, \ldots n_L}) \in \bigotimes^L \CZ^N| ~ ~ |n_{L+1} \rangle \rangle = \omega^{-{\tt r}n_1} | n_1 \rangle \rangle , ~ ~ {\tt Z} (w) = \omega^{\tt Q} w \},  
$$
and hence follows the isomorphism: 
\be
\Theta : V_{r, Q} \cong W_{{\tt r}, {\tt Q}} , ~ ~ |Q; n_1, \ldots n_L \rangle \mapsto |n_1, \ldots n_L \rangle \rangle, ~ ~ ({\tt r}, {\tt Q}) = (Q, r).
\ele(Theta)
Using (\req(taupd)), (\req(Uf)), (\req(FaLpp')) and {\req(Ftau2)), one obtains
$$
\langle Q; n_1, \ldots n_L | \tau^{(2)}(t; p,p') |Q; n'_1, \ldots n'_L \rangle = \langle \langle n_1, \ldots n_L | \tau_F^{(2)}(t; p, p')|n'_1, \ldots n'_L \rangle \rangle
$$
where $\sum_\ell n_\ell \equiv  \sum_\ell n'_\ell \equiv r$, and $\tau_F^{(2)}(t; p, p')$ with the boundary condition  ${\tt r}\equiv Q$. Equivalently, $\tau^{(2)}(t; p,p')$ in the charge-$Q$ sector with boundary condition $r$  is equivalent to $\tau_F^{(2)}(t; p, p')$ in the charge-${\tt Q} ~(=r)$ sector with boundary condition ${\tt r}~( = Q)$: 
\be
 \tau^{(2)}(t; p,p')  = \Theta^{-1} \tau_F^{(2)}(t; p, p')  \Theta .
\ele(ttf)
Combining (\req(tf=t)) and  (\req(ttf)), we have shown the following identical $\tau^{(2)}$-models:
\begin{prop}\label{prop:t=t*}
For rapidities $p, p'$ in a chiral-Potts curve ${\goth W}$ in $(\req(cpmC))$, let $p^*, p'^*$ be the dual rapidities in the ${\goth W}^*$ defined by $(\req(W*))$. Then $\tau^{(2)}(t; p,p')$ in the charge $Q$ sector with  boundary condition $r$  is equivalent to  $\tau^{(2)}(t^*; {p'}^*,p^*)$ in the charge $Q^*$ sector with boundary condition $r^*$ when $(Q^*, r^*) =(r, Q)$ . Indeed, $\tau^{(2)}(t; p,p')$ on $V_{r, Q}$ is similar to $\tau^{(2)}(t^*; p'^*,p^*)$ on $V_{r^*, Q^*}$:
$$
\tau^{(2)}(t; p,p')  = \Psi^{-1} \tau^{(2)}(t^*; p'^*, p^*)  \Psi , ~ ~ ~ t^*= (-1)^\frac{1}{N} t ,
$$
where $\Psi$ is the isomorphism, 
\be
\Psi : V_{r, Q} \longrightarrow V_{r^*, Q^*}, ~ ~ |Q; n_1, \ldots n_L \rangle \mapsto |\widehat{n_1}, \ldots \widehat{n_L} \rangle, ~ ~ (\sum_{\ell=1}^L n_\ell \equiv r)  
\ele(Psi)
with $(Q^*, r^*) =(r, Q)$.
\end{prop}
{\bf Remark.} There are two bases for the vector space $V_{r, Q}$ in  (\req(VrQ)):
$$
\begin{array}{lll}
V_{r, Q} & = \bigoplus_{n_\ell} \CZ |Q; n_1, \ldots n_L \rangle & (\sum_{\ell=1}^L n_\ell \equiv r, ~ ~ n_{L+1} = \omega^{-Qn_1} n_1) \\
&  = \bigoplus_{n'_\ell } \CZ |\widehat{n'_1}, \ldots \widehat{n'_L} \rangle & (\sum_{\ell=1}^L n'_\ell \equiv Q , ~ ~ |\widehat{n'_{L+1}}\rangle = \omega^{-rn'_1} |\widehat{n'_1}\rangle ),
\end{array}
$$
which are related by
$$
\begin{array}{ll}
|\widehat{n'_1}, \ldots \widehat{n'_L} \rangle 
&= N^\frac{-(L-1)}{2} \sum_{n_\ell } \omega^{ \sum_{\ell=1}^L n'_\ell (n_1+\ldots +n_{\ell-1}) } | Q; n_1, \ldots n_L \rangle ; \\
| Q; n_1, \ldots n_L \rangle &
= N^\frac{-(L-1)}{2} \sum_{n'_\ell} \omega^{-\sum_\ell (n_1+\ldots+n_{\ell-1}) n'_\ell}   |\widehat{n'_1}, \ldots \widehat{n'_L}  \rangle .
\end{array}
$$
Since the spatial translation $S_R$ acts on $V_{r, Q}$  by
$$
S_R| Q; n_1, \ldots n_L \rangle = \omega^{- Q n_1 }|Q; n_2, \ldots n_L, n_{L+1} \rangle , ~ ~ ~ ~
S_R|\widehat{n'_1}, \ldots \widehat{n'_L} \rangle 
=|\widehat{n'_2}, \ldots \widehat{n'_L}, \widehat{n'_{L+1}} \rangle ,
$$
the isomorphism $\Psi$ in (\req(Psi)) is $S_R$-equivariant, also described by
$$
\Psi : V_{r, Q} \longrightarrow V_{r^*, Q^*}, ~ ~ |\widehat{n'_1}, \ldots \widehat{n'_L} \rangle \mapsto 
\omega^{ (Q-n'_1)Q^* }  | Q^*; n'_2, \ldots n'_L, n'_1 \rangle .
$$

\subsection{Duality relation in chiral Potts models  \label{ssec.CPCP*}}
We now extend the duality of $\tau^{(2)}$-model to CPM. First we note that there exists a $k' \leftrightarrow k'^{-1}$ duality about the Boltzmann weights in (\req(Weig)). Indeed, for rapidities $p, q$ in a chiral Potts curve ${\goth W}$ in (\req(cpmC)), the Fourier-forms of Boltzmann weights (\req(WWf)) equal to Boltzmann weights for the dual rapidities $p^*, q^* \in {\goth W}^*$:
\bea(ll)
\frac{\overline{W}^{(f)}_{p q }(k)}{\overline{W}^{(f)}_{pq}(0)} = W_{p^*q^*}(k)  ,  & \frac{W^{(f)}_{p q}(k)}{W^{(f)}_{p q}(0)} =  \overline{W}_{p^*q^*}(-k), ~ ~ (k \in \ZZ_N ).
\elea(WW*)
Note that when $k'= \pm 1$, ${\goth W}^*$ can be identified with ${\goth W}$ via the isomorphism, $({\rm i}^\frac{1}{N} x^*, {\rm i}^\frac{1}{N} y^*, \mu^*) \mapsto ( x^*,  y^*, \mu^*)$, preserving Boltzmann weights. The above relation describes the self-dual property of the Boltzmann weights at $k'= \pm 1$, in particular for the case ${\goth W}= \overline{\goth W}_{\pm 1}$ in  \cite{FatZ}. Using (\req(WW*)), one finds the following expression of the chiral-Potts-transfer matrix (\req(ThatT)) for $\in {\goth W}^*$, 
\bea(l)
\langle \widehat{n_1}, \ldots \widehat{n_L}|T (q^*; p'^*, p^*) |\widehat{n'_1}, \ldots \widehat{n'_L} \rangle = 
(N \overline{W}^{(f)}_{p' ,q}(0 ) W^{(f)}_{p ,q}(0))^{-L}  \\
\times \sum_{k_\ell, k'_\ell} \omega^{\sum_{\ell=1}^L (n_\ell k_\ell - \sum_\ell n'_\ell k'_\ell)}
\prod_{\ell =1}^L \overline{W}^{(f)}_{p' q}(k_\ell - k'_\ell ) W^{(f)}_{p q}(- k_{\ell+1} + k'_\ell), \\
\langle \widehat{n_1}, \ldots \widehat{n_L}|\widehat{T} (q^*; p'^*, p^*) |\widehat{n'_1}, \ldots \widehat{n'_L} \rangle = (N W^{(f)}_{p' ,q}(0 ) \overline{W}^{(f)}_{p ,q}(0))^{-L} \\
\times \sum_{k_\ell, k'_\ell} \omega^{  \sum_\ell (n_\ell k_\ell -  n'_\ell k'_\ell) }
\prod_{\ell =1}^L W^{(f)}_{p' q}(-k_\ell + k'_\ell ) \overline{W}^{(f)}_{p q}(k_\ell - k'_{\ell+1}),
\elea(Tp*)
where the indices $k_\ell, k_\ell's$ are in $\ZZ_N$ with the boundary condition, $k_{L+1} = k_1 - r^* , k_{L+1}'= k_1'- r^*$. On the other hand, when $\sum_{\ell =1}^L n_\ell \equiv \sum_{\ell =1}^L n_\ell \equiv r$, one has
$$
\begin{array}{l}
\langle Q; n_1, \ldots n_L|T (q; p, p') |Q; n'_1, \ldots n'_L \rangle =  
\frac{1}{N} \sum_{\sigma_1, \sigma'_1=0}^{N-1}\omega^{Q (\sigma_1-\sigma'_1)} \prod_{\ell =1}^L W_{pq}(\sigma_\ell - \sigma'_\ell ) \overline{W}_{p' q}(\sigma_{\ell+1} - \sigma'_\ell) \\
= \frac{1}{N^{L+1}} 
\sum_{\sigma_1, \sigma'_1, m_\ell, m'_\ell } \omega^{Q (\sigma_1-\sigma'_1)- \sum_{\ell=1}^L \{ (\sigma_\ell - \sigma'_\ell )m_\ell + (\sigma_{\ell+1} - \sigma'_\ell) m'_\ell \} }\prod_{\ell =1}^L  W^{(f)}_{pq}(m_\ell ) \overline{W}^{(f)}_{p' q}(m'_\ell)  \\
= \frac{1}{N^{L-1}} 
\sum_{ m'_\ell } \omega^{\sum_\ell n'_\ell  m_\ell' } (\prod_\ell \overline{W}^{(f)}_{p' q}(m'_\ell)) 
(\sum_{m_\ell}^\prime \omega^{\sum_\ell (-n_1- \ldots -n_{\ell-1}+n'_1+ \ldots +n'_{\ell-1} ) (m_\ell - m'_{\ell-1} ) } \prod_{\ell =1}^L  W^{(f)}_{pq}(- m_\ell ) ) ,
\end{array}
$$
where all indices are in $\ZZ_N$ with $\sigma_{\ell+1} = \sigma_\ell - n_\ell$,  $\sigma'_{\ell+1} = \sigma'_\ell - n'_\ell$, $m'_L=m'_0$, and the prime-summation $\sum_{m_\ell}^\prime$ means the indices $m_\ell$s satisfying $\sum_\ell m_\ell \equiv -Q + \sum_\ell m_\ell'$. The above formula, up to a scale factor, can be identified with the first expression in (\req(Tp*)) under the constraint $r^*= Q$ by the change of indices $k_\ell - k_\ell' = m_\ell', k_\ell - k_{\ell- 1} = - m_\ell + m_{\ell-1}'$. Similarly, the same relation between $\widehat{T} (q; p, p')$ and $\widehat{T} (q^*; p'^*, p^*)$ holds. Indeed, one finds
$$
\begin{array}{l}
\langle Q; n_1, \ldots n_L|T (q; p, p') |Q; n'_1, \ldots n'_L \rangle =(\overline{W}^{(f)}_{p' ,q}(0 ) W^{(f)}_{p ,q}(0))^L\langle \widehat{n_1}, \ldots \widehat{n_L}|T (q^*; p'^*, p^*) |\widehat{n'_1}, \ldots \widehat{n'_L} \rangle , \\
\langle Q; n_1, \ldots n_L|\widehat{T} (q; p, p') |Q; n'_1, \ldots n'_L \rangle =(W^{(f)}_{p' ,q}(0 ) \overline{W}^{(f)}_{p ,q}(0))^L\langle \widehat{n_1}, \ldots \widehat{n_L}|\widehat{T}(q^*; p'^*, p^*) |\widehat{n'_1}, \ldots \widehat{n'_L} \rangle .
\end{array}  
$$
Using $W^{(f)}_{p^* ,q^*}(0)= \overline{W}^{(f)}_{p ,q }(0)^{-1}$,
we have shown the following duality relation in CPM as an extention of the $\tau^{(2)}$-duality in Proposition \ref{prop:t=t*}.
\begin{thm}\label{thm:dualCP}
For a $k'$-curve ${\goth W}$ in $(\req(cpmC))$ for $k' \neq 0$, let ${\goth W}^*$ be the dual $k'^{-1}$-curve defined in $(\req(W*))$. Then the chiral Potts models with rapidities in ${\goth W}$ and ${\goth W}^*$ are equivalent when the $\ZZ_N$-charge and the boundary condition are interchanged.  More precisely, the chiral-Potts transfer matrices over ${\goth W}$ in the $Q$ sector with the boundary condition $r$ are similar to those over ${\goth W}^*$ in the $Q^*$ sector with the boundary condition $r^*$ via the isomorphism $(\req(Psi))$ when $(Q^*, r^*) =(r, Q)$: 
\bea(l)
W^{(f)}_{p ,q}(0)^{-L} T (q; p, p') = W^{(f)}_{p'^* ,q^*}(0)^{-L} \Psi^{-1} T (q^*; p'^*, p^*) \Psi , \\
W^{(f)}_{p' ,q}(0)^{-L} \widehat{T} (q; p, p') = W^{(f)}_{p^* ,q^*}(0)^{-L} \Psi^{-1}\widehat{T}(q^*; p'^*, p^*)  \Psi 
\elea(TT*)
where $\Psi$ is the isomorphism in $(\req(Psi))$ and $p, p', q \in {\goth W}$ with the corresponding dual $p^*, p'^*, q^* \in {\goth W}^*$.
\end{thm}
{\bf Remark.} By the same argument, Theorem \ref{thm:dualCP} is also valid for inhomogeneous CPM with vertical rapidities $\{ p_1,\ldots p_L\}, \{p'_1, \ldots p'_L\}$. The relation (\req(TT*)) still holds, but with one shifting the indices of ${p_\ell}^*$s  by one, i.e. the operators $T (q^*;\{p_1'^*,\ldots p_L'^*\},\{p_2^*,\ldots p_{L+1}^*\})$, $\widehat{T} (q^*;\{p_1'^*,\ldots p_L'^*\},\{p_2^*,\ldots p_{L+1}^*\})$; a similar statement also for the $\tau^{(2)}$-model in Proposition \ref{prop:t=t*}.

\subsection{Chiral Potts model of $\tau^{(2)}$-face-model, and chiral Potts model of the dual lattice  \label{ssec.CP*}}
In this subsection, we form the chiral Potts model over the dual lattice, as well as that of $\tau^{(2)}$-face-model, as in \cite{B89} section 5 for the superintegrable case. 
Introduce the ${\tt C}^N$-basis,
$$
|\sigma^* \rangle^* = \frac{1}{\sqrt{N}} \sum_{n=0}^{N-1} \omega^{\sigma^* n} | n \rangle \rangle \in {\tt C}^N, ~ ~ \sigma^*  \in \ZZ_N, 
$$
by which $\Phi$ in (\req(Phi)) is defined by $|\sigma^* \rangle^* \mapsto |\sigma^*  \rangle$. As in (\req(XZF)), the ${\tt C}^N$-Weyl-operators $({\tt X}, {\tt Z}) $ can be expressed by the basis  $|\sigma^* \rangle^*$ with 
$({\tt X}^*, {\tt Z}^*) = ({\tt Z}, {\tt X}^{-1})$,  where ${\tt X}^*|\sigma^* \rangle^* = |\sigma^*+1 \rangle^*$ and ${\tt Z}^*|\sigma^* \rangle^*= \omega^{\sigma^*} |\sigma^* \rangle^*$.
Using the ${\tt C}^N$-basis $|\sigma^* \rangle^*$, one defines the chiral Potts model over the dual lattice as follows. The chiral Potts model $T (q ; p, p')$ or $\widehat{T} (q ; p, p')$ in (\req(ThatT))
with rapidities in a curve ${\goth W}$ in (\req(cpmC)) is defined on a square lattice $\Gamma$ by attaching a $\CZ^N$-quantum space over each vertex, (see Figure 1 with $\circ$ as vertices and $\star$ as faces).

The dual lattice $\Gamma^*$ of $\Gamma$ is a square lattice whose vertices (or faces) are in one-to-one correspondence with the faces (vertices respectively) of $\Gamma$. 
Using the same Boltzmann weights (\req(Weig)) with rapidities in ${\goth W}$, the chiral Potts model, ${\tt T}^* (q ; p, p'), \widehat{\tt T}^* (q ; p, p')$, is defined on the dual lattice $\Gamma^*$ by attaching a ${\tt C}^N$-quantum space on each vertex of $\Gamma^*$.  The transfer matrices are defined by
\bea(l)
{\tt T}^* (q)_{\{\sigma^* \}, \{\sigma'^* \}} (= {\tt T}^* (q ; { \{p_\ell \}, \{ p'_\ell \} })_{\{\sigma^* \}, \{\sigma'^* \}})= \prod_{\ell =1}^L \overline{W}_{p_\ell q}(\sigma^*_\ell - \sigma^{* ~ \prime}_{\ell-1} )W_{p'_\ell q}(\sigma^*_\ell - \sigma^{*  \prime}_\ell), \\
\widehat{\tt T}^* (q)_{\{\sigma^{*  \prime} \}, \{\sigma^{* \prime \prime}\}}(=\widehat{\tt T}^*(q ; { \{p_\ell \}, \{ p'_\ell \} } )_{\{\sigma^{*  \prime} \}, \{\sigma^{* \prime \prime} \}})= \prod_{\ell =1}^L W_{p_\ell q}(\sigma^{*  \prime}_{\ell-1} - \sigma^{* \prime \prime}_\ell) \overline{W}_{p'_\ell q}(\sigma^{* \prime}_\ell - \sigma^{* \prime \prime}_\ell) ,
\elea(ThatT*)
(see Figure 2 about ${\tt T}^*(q^*; p^*, p^{* \prime}), \widehat{\tt T}^*(q^*; p^*, p^{* \prime})$). By (\req(ThatT)) and (\req(ThatT*)), ${\tt T}^*(q; p, p'), \widehat{\tt T}^*(q; p, p')$ are the same as $T (q; p', p), \widehat{T}(q; p', p)$ respectively by identifying ${\tt C}^N$ with $\CZ^N$ via $\Phi$ in (\req(Phi)):
\be
{\tt T}^*(q; p, p') = \Phi^{-1} T (q; p', p) \Phi , ~ ~ \widehat{\tt T}^*(q; p, p') = \Phi^{-1} \widehat{T}(q; p', p) \Phi \ .
\ele(T*=T)
Theorem \ref{thm:dualCP} can be stated as the relation between a chiral Potts model in ${\goth W}$-rapidities on $\Gamma$-lattice (in the $Q$-sector and $r$-boundary condition) and  the dual chiral Potts model in ${\goth W}^*$-rapidities on $\Gamma^*$-lattice (in the ${\tt Q}^*$-sector and ${\tt r}^*$-boundary condition) for $(Q, r) = ({\tt r}^*, {\tt Q}^*)$:
\bea(l)
W^{(f)}_{p ,q}(0)^{-L}T (q; p, p')  = W^{(f)}_{p'^* ,q^*}(0)^{-L}  \Theta^{-1} {\tt T}^* (q^*; p^*, p'^* ) \Theta , \\  W^{(f)}_{p' ,q}(0)^{-L} \widehat{T} (q; p, p')  = W^{(f)}_{p^* ,q^*}(0)^{-L}  \Theta^{-1} \widehat{\tt T}^* (q^*;  p^*, p'^*)  \Theta
\elea(TT*l)
where $\Theta$ is the isomorphism defined in (\req(Theta)).  Note that the above formula is still valid in the inhomogeneous case by changing $(p, p'), (p^*, p'^*)$ to $(\{p_\ell \}, \{ p'_\ell \}), (\{p^*_\ell \}, \{p^{\prime *}_\ell \})$ respectively.

$$
\put (20 , -100 ){\vector( 0, 1){200}}
\put (20 , -110 ){\shortstack{ $p$ }}
\put (60 , -100 ){\vector( 0, 1){200}}
\put (60 , -110 ){\shortstack{ $p'$ }}
\put (162 , 20 ){\vector( -1, 0){350}}
\put (-200 , 18 ){\shortstack{ $q$ }}
\put (0 , -90 ){\shortstack{ $\sigma_\ell$ }}
\put (40 , -30 ){\shortstack{ $\sigma'_\ell$ }}
\put (-160 , -90 ){\shortstack{ $\sigma_1$ }}
\put (72 , -90 ){\shortstack{ $\cdots \sigma_L$ }}
\put (160 , -90 ){\shortstack{ $\sigma_{L+1}$ }}
\put ( 0, 22 ){\shortstack{ $W_{pq}$ }}
\put ( 65, 24 ){\shortstack{ $\overline{W}_{p'q}$ }}
\put ( 160, 75 ){\shortstack{ $\circ$ }}
\put ( 160, 35 ){\shortstack{ $\star$ }}
\put ( 160, -5 ){\shortstack{ $\circ$  }}
\put ( 160, -45 ){\shortstack{ $\star$  }}
\put ( 160, -85 ){\shortstack{ $\circ$  }}
\put ( 160, -1 ){\line(-1,1){40}}
\put ( 160, -81 ){\line(-1,1){40}}
\put ( 120, 75 ){\shortstack{ $\star$ }}
\put ( 120, 35 ){\shortstack{$\circ$ }}
\put ( 120, -5 ){\shortstack{$\star$ }}
\put ( 120, -45 ){\shortstack{ $\circ$ }}
\put ( 120, -85 ){\shortstack{ $\star$  }}
\put ( 120, 37 ){\line(1, 1){40}}
\put ( 120, 39 ){\line(-1,1){40}}
\put ( 120, -43 ){\line(1, 1){40}}
\put ( 120, -41 ){\line(-1,1){40}}
\put ( 80, 75 ){\shortstack{ $\circ$ }}
\put ( 80, 35 ){\shortstack{ $\star$ }}
\put ( 80, -5 ){\shortstack{ $\circ$  }}
\put ( 80, -45 ){\shortstack{ $\star$  }}
\put ( 80, -85 ){\shortstack{ $\circ$  }}
\put ( 80, -3 ){\line(1, 1){40}}
\put ( 80, -1 ){\vector(-1,1){36}}
\put ( 80, -83 ){\line(1, 1){40}}
\put ( 80, -81 ){\line(-1,1){40}}
\put ( 40, 75 ){\shortstack{ $\star$ }}
\put ( 40, 35 ){\shortstack{$\circ$ }}
\put ( 40, -5 ){\shortstack{$\star$ }}
\put ( 40, -45 ){\shortstack{ $\circ$ }}
\put ( 40, -85 ){\shortstack{ $\star$  }}
\put ( 40, 37 ){\line(1, 1){40}}
\put ( 40, 39 ){\line(-1,1){40}}
\put ( 40, -43 ){\line(1, 1){40}}
\put ( 40, -41 ){\line(-1,1){40}}
\put ( 0, 75 ){\shortstack{ $\circ$ }}
\put ( 0, 35 ){\shortstack{ $\star$ }}
\put ( 0, -5 ){\shortstack{ $\circ$  }}
\put ( 0, -45 ){\shortstack{ $\star$  }}
\put ( 0, -85 ){\shortstack{ $\circ$  }}
\put ( 0, -3 ){\vector(1, 1){38}}
\put ( 0, -1 ){\line(-1,1){40}}
\put ( 0, -83 ){\line(1, 1){40}}
\put ( 0, -81 ){\line(-1,1){40}}
\put ( -40, 75 ){\shortstack{ $\star$ }}
\put ( -40, 35 ){\shortstack{ $\circ$}}
\put ( -40, -5 ){\shortstack{ $\star$  }}
\put ( -40, -45 ){\shortstack{ $\circ$ }}
\put ( -40, -85 ){\shortstack{ $\star$ }}
\put ( -40, 37 ){\line(1, 1){40}}
\put ( -40, 39 ){\line(-1,1){40}}
\put ( -40, -43 ){\line(1, 1){40}}
\put ( -40, -41 ){\line(-1,1){40}}
\put ( -80, 75 ){\shortstack{ $\circ$ }}
\put ( -80, 35 ){\shortstack{ $\star$ }}
\put ( -80, -5 ){\shortstack{ $\circ$  }}
\put ( -80, -45 ){\shortstack{ $\star$  }}
\put ( -80, -85 ){\shortstack{ $\circ$  }}
\put ( -80, -3 ){\line(1, 1){40}}
\put ( -80, -1 ){\line(-1,1){40}}
\put ( -80, -83 ){\line(1, 1){40}}
\put ( -80, -81 ){\line(-1,1){40}}
\put ( -120, 75 ){\shortstack{ $\star$ }}
\put ( -120, 35 ){\shortstack{ $\circ$}}
\put ( -120, -5 ){\shortstack{ $\star$  }}
\put ( -120, -45 ){\shortstack{ $\circ$ }}
\put ( -120, -85 ){\shortstack{ $\star$ }}
\put ( -120, 37 ){\line(1, 1){40}}
\put ( -120, 39 ){\line(-1,1){40}}
\put ( -120, -43 ){\line(1, 1){40}}
\put ( -120, -41 ){\line(-1,1){40}}
\put ( -160, 75 ){\shortstack{ $\circ$ }}
\put ( -160, 35 ){\shortstack{ $\star$ }}
\put ( -160, -5 ){\shortstack{ $\circ$  }}
\put ( -160, -45 ){\shortstack{ $\star$  }}
\put ( -160, -85 ){\shortstack{ $\circ$  }}
\put ( -160, -3 ){\line(1, 1){40}}
\put ( -160, -83 ){\line(1, 1){40}}
\put ( -200, -150 ){\shortstack{Figure 1. The square lattice $\Gamma$ for the chiral Potts model with the skewed boundary\\ condition $\sigma_{L+1}=\sigma_1-r$, showing rapidities of two vertical $p,p'$ and one horzontal $q$.} }
$$
\par \vspace{2mm} \noindent
$$
\put (20 , -100 ){\vector( 0, 1){200}}
\put (20 , -110 ){\shortstack{ $p^*$ }}
\put (60 , -100 ){\vector( 0, 1){200}}
\put (60 , -110 ){\shortstack{ $p^{* \prime}$ }}
\put (162 , 20 ){\vector( -1, 0){350}}
\put (-200 , 18 ){\shortstack{ $q^*$ }}
\put (40 , -90 ){\shortstack{ $\sigma^*_\ell$ }}
\put (80 , -30 ){\shortstack{ ${\sigma_\ell^*}'$ }}
\put (-120 , -90 ){\shortstack{ $\sigma^*_1$ }}
\put (112 , -90 ){\shortstack{ $\cdots \sigma^*_L$ }}
\put (0 , -90 ){\shortstack{ $\sigma_\ell$ }}
\put (40 , -32 ){\shortstack{ $\sigma'_\ell$ }}
\put ( 20, 20 ){\shortstack{ $\overline{W}_{p^*q^*}$ }}
\put ( 64, 12 ){\shortstack{ $W_{p'^*q^*}$ }}
\put ( 160, 75 ){\shortstack{ $\circ$ }}
\put ( 160, 35 ){\shortstack{ $\star$ }}
\put ( 160, -5 ){\shortstack{ $\circ$  }}
\put ( 160, -45 ){\shortstack{ $\star$  }}
\put ( 160, -85 ){\shortstack{ $\circ$  }}
\put ( 162, 39 ){\line(-1, 1){40}}
\put ( 162, 37 ){\line(-1, 1){40}}
\put ( 162, -41 ){\line(-1, 1){40}}
\put ( 162, -43 ){\line(-1, 1){40}}
\put ( 120, 75 ){\shortstack{ $\star$ }}
\put ( 120, 35 ){\shortstack{$\circ$ }}
\put ( 120, -5 ){\shortstack{$\star$ }}
\put ( 120, -45 ){\shortstack{ $\circ$ }}
\put ( 120, -85 ){\shortstack{ $\star$  }}
\put ( 123, -1 ){\line(1, 1){40}}
\put ( 124, -2 ){\line(1, 1){40}}
\put ( 122, -1 ){\line(-1, 1){40}}
\put ( 122, -3 ){\line(-1, 1){40}}
\put ( 123, -81 ){\line(1, 1){40}}
\put ( 124, -82 ){\line(1, 1){40}}
\put ( 122, -81 ){\line(-1, 1){40}}
\put ( 122, -83 ){\line(-1, 1){40}}
\put ( 80, 75 ){\shortstack{ $\circ$ }}
\put ( 80, 35 ){\shortstack{ $\star$ }}
\put ( 80, -5 ){\shortstack{ $\circ$  }}
\put ( 80, -45 ){\shortstack{ $\star$  }}
\put ( 80, -85 ){\shortstack{ $\circ$  }}
\put ( 83, 39 ){\line(1, 1){40}}
\put ( 84, 38 ){\line(1, 1){40}}
\put ( 82, 39 ){\line(-1, 1){40}}
\put ( 82, 37 ){\line(-1, 1){40}}
\put ( 83, -41 ){\line(1, 1){40}}
\put ( 84, -42 ){\line(1, 1){40}}
\put ( 82, -41 ){\line(-1, 1){40}}
\put ( 82, -43 ){\line(-1, 1){40}}
\put ( 40, 75 ){\shortstack{ $\star$ }}
\put ( 40, 35 ){\shortstack{$\circ$ }}
\put ( 40, -5 ){\shortstack{$\star$ }}
\put ( 40, -45 ){\shortstack{ $\circ$ }}
\put ( 40, -85 ){\shortstack{ $\star$  }}
\put ( 43, -1 ){\line(1, 1){40}}
\put ( 44, -2 ){\line(1, 1){40}}
\put ( 42, -1 ){\line(-1, 1){40}}
\put ( 42, -3 ){\line(-1, 1){40}}
\put ( 43, -81 ){\line(1, 1){40}}
\put ( 44, -82 ){\line(1, 1){40}}
\put ( 42, -81 ){\line(-1, 1){40}}
\put ( 42, -83 ){\line(-1, 1){40}}
\put ( 0, 75 ){\shortstack{ $\circ$ }}
\put ( 0, 35 ){\shortstack{ $\star$ }}
\put ( 0, -5 ){\shortstack{ $\circ$  }}
\put ( 0, -45 ){\shortstack{ $\star$  }}
\put ( 0, -85 ){\shortstack{ $\circ$  }}
\put ( 3, 39 ){\line(1, 1){40}}
\put ( 4, 38 ){\line(1, 1){40}}
\put ( 2, 39 ){\line(-1, 1){40}}
\put ( 2, 37 ){\line(-1, 1){40}}
\put ( 3, -41 ){\line(1, 1){40}}
\put ( 4, -42 ){\line(1, 1){40}}
\put ( 2, -41 ){\line(-1, 1){40}}
\put ( 2, -43 ){\line(-1, 1){40}}
\put ( -40, 75 ){\shortstack{ $\star$ }}
\put ( -40, 35 ){\shortstack{ $\circ$}}
\put ( -40, -5 ){\shortstack{ $\star$  }}
\put ( -40, -45 ){\shortstack{ $\circ$ }}
\put ( -40, -85 ){\shortstack{ $\star$ }}
\put ( -37, -1 ){\line(1, 1){40}}
\put ( -36, -2 ){\line(1, 1){40}}
\put ( -38, -1 ){\line(-1, 1){40}}
\put ( -38, -3 ){\line(-1, 1){40}}
\put ( -37, -81 ){\line(1, 1){40}}
\put ( -36, -82 ){\line(1, 1){40}}
\put ( -38, -81 ){\line(-1, 1){40}}
\put ( -38, -83 ){\line(-1, 1){40}}
\put ( -80, 75 ){\shortstack{ $\circ$ }}
\put ( -80, 35 ){\shortstack{ $\star$ }}
\put ( -80, -5 ){\shortstack{ $\circ$  }}
\put ( -80, -45 ){\shortstack{ $\star$  }}
\put ( -80, -85 ){\shortstack{ $\circ$  }}
\put ( -77, 39 ){\line(1, 1){40}}
\put ( -76, 38 ){\line(1, 1){40}}
\put ( -78, 39 ){\line(-1, 1){40}}
\put ( -78, 37 ){\line(-1, 1){40}}
\put ( -77, -41 ){\line(1, 1){40}}
\put ( -76, -42 ){\line(1, 1){40}}
\put ( -78, -41 ){\line(-1, 1){40}}
\put ( -78, -43 ){\line(-1, 1){40}}
\put ( -120, 75 ){\shortstack{ $\star$ }}
\put ( -120, 35 ){\shortstack{ $\circ$}}
\put ( -120, -5 ){\shortstack{ $\star$  }}
\put ( -120, -45 ){\shortstack{ $\circ$ }}
\put ( -120, -85 ){\shortstack{ $\star$ }}
\put ( -117, -1 ){\line(1, 1){40}}
\put ( -116, -2 ){\line(1, 1){40}}
\put ( -118, -1 ){\line(-1, 1){40}}
\put ( -118, -3 ){\line(-1, 1){40}}
\put ( -117, -81 ){\line(1, 1){40}}
\put ( -116, -82 ){\line(1, 1){40}}
\put ( -118, -81 ){\line(-1, 1){40}}
\put ( -118, -83 ){\line(-1, 1){40}}
\put ( -160, 75 ){\shortstack{ $\circ$ }}
\put ( -160, 35 ){\shortstack{ $\star$ }}
\put ( -160, -5 ){\shortstack{ $\circ$  }}
\put ( -160, -45 ){\shortstack{ $\star$  }}
\put ( -160, -85 ){\shortstack{ $\circ$  }}
\put ( -157, 39 ){\line(1, 1){40}}
\put ( -156, 38 ){\line(1, 1){40}}
\put ( -157, -41 ){\line(1, 1){40}}
\put ( -156, -42 ){\line(1, 1){40}}
\put ( -200, -150 ){\shortstack{Figure 2. The dual square lattice $\Gamma^*$ for the chiral Potts model with the skewed boundary\\ condition $\sigma^*_{L+1}=\sigma^*_1-r^*$, showing rapidities of two vertical $p^*, {p'}^*$ and one horzontal $q^*$.} }
$$
\par \vspace{1mm}

We now define the chiral Potts model, $T_F(q, p, p')$ or $\widehat{T}_F(q, p, p')$, over $\Gamma^*$-lattice for the $\tau^{(2)}$-face-model $\tau_F^{(2)}(t; p, p')$ in (\req(Ftau2)). As before, the ${\tt L}$-operator (\req(FaLpp')) of $\tau_F^{(2)}(t; p, p')$ for $p, p' \in {\goth W}$ is equivalent to the $L$-operator (\req(L)) of $\tau^{(2)}(t^*; p'^*, p^*)$ for the dual rapidities $p^*, p'^*  \in {\goth W}^*$ via the isomorphism $\Phi$ in (\req(Phi)). Therefore, the chiral Potts model $T_F(q, p, p'), \widehat{T}_F(q, p, p')$ with $p, p', q \in {\goth W}$ is identified with the CPM over the dual curve ${\goth W}^*$ via the dual map (\req(pp*)) so that the following relations hold,
$$
T_F(q, p, p')= \Phi^{-1} T (q^*; p'^*, p^*)  \Phi , ~ ~ \widehat{T}_F(q, p, p')= \Phi^{-1} \widehat{T} (q^*; p'^*, p^*)  \Phi ,
$$
extending that in (\req(tf=t)). By (\req(T*=T)), the above relations are the same as
$$
T_F(q, p, p')= {\tt T}^* (q^*; p^*, p'^*) , ~ ~ \widehat{T}_F(q, p, p')=  \widehat{\tt T}^* (q^*; p^*, p'^*)  ,
$$
with the identical charge and boundary condition, $({\tt Q}, {\tt r})= ({\tt Q}^*, {\tt r}^*)$. By the duality relation (\req(TT*l)), one obtains the extended relation of (\req(ttf)):
$$
\begin{array}{l}
W^{(f)}_{p ,q}(0)^{-L} T (q; p, p')  = W^{(f)}_{p'^* ,q^*}(0)^{-L}  \Theta^{-1} T_F(q, p, p') \Theta , \\  W^{(f)}_{p' ,q}(0)^{-L} \widehat{T} (q; p, p')  = W^{(f)}_{p^* ,q^*}(0)^{-L}   \Theta^{-1} T_F(q, p, p')  \Theta ,
\end{array}
$$
with the constraint, $(Q, r)= ({\tt r}, {\tt Q})$.

\subsection{Kramers-Wannier duality in Ising model \label{ssec.Ising}}
For CPM in $N=2$ case, 
we regain the Ising model with the following homogeneous parameterization of rapidities $p \in {\goth W}_{k'}$ for $k' \neq 0, \pm 1$ (\cite{B93c} section 3),
$$ 
\begin{array}{l}
a_p: b_p : c_p : d_p = - H(u_p) : - H_1 (u_p) : \Theta_1 (u_p) : \Theta (u_p) , 
\end{array}
$$
equivalently the coordinates of ${\goth W}_{k'}$ in (\req(cpmC)) given by
$$
x_p = - k^\frac{1}{2} {\rm sn} u_p , ~ ~ y_p = - k^\frac{1}{2} \frac{{\rm cn} u_p}{{\rm dn} u_p}, ~ ~ \mu_p =  {k'}^\frac{1}{2} \frac{1}{{\rm dn}  u_p} .    
$$
Here the Jacobi theta functions are of modulus $k$ with the elliptical integrals $(K , K') = (K (k), K'(k)) = (K'(k'), K (k'))$. The Boltzmann weights (\req(Weig)) are now expressed by 
$$ 
\begin{array}{lll}
W_{pq}(0)=1, & W_{pq}(1)= \frac{{\rm cn} u_q + {\rm sn} u_p {\rm dn} u_q }{{\rm cn} u_p + {\rm sn} u_q {\rm dn} u_p } &= k' {\rm scd} (u_p- u_q + K) ; \\
\overline{W}_{pq}(0)=1 , & \overline{W}_{pq}(1) = \frac{k'( {\rm sn} u_q - {\rm sn} u_p)}{{\rm dn} u_p {\rm cn} u_q + {\rm dn} u_q {\rm cn} u_p} &= k' {\rm scd} (u_q- u_p )
\end{array}
$$
where ${\rm scd} (u ):= \frac{{\rm sn} (u/2) }{{\rm cn} (u/2) {\rm dn} (u/2)} = \frac{{\rm sn} u}{{\rm cn} u + {\rm dn} u}$. If $J$ and $\overline{J}$ are the usual dimensionless Ising model intersection coefficients, $W_{pq}(1)= {\rm exp}(-2J), \overline{W}_{pq}(1) = {\rm exp}(-2\overline{J})$, and we find \cite{Bax} (7.8.5),
$$
\sinh 2J = \frac{{\rm sn} u}{{\rm cn} u} , ~ ~ \sinh 2\overline{J} = \frac{{\rm cn} u}{k' {\rm sn} u}  ~ ~ ~ (u= u_q - u_p) .
$$
Similarly, the coordinates of the rapidity $p^*$ in ${\goth W}_\frac{1}{k'}$ are expressed by
$$
x_{p^*} = - (\frac{{\rm i} k}{k'})^\frac{1}{2} {\rm sn}^* u_{p^*} , ~ ~ y_{p^*} = - (\frac{{\rm i} k}{k'})^\frac{1}{2} \frac{{\rm cn}^* u_{p^*}}{{\rm dn}^* u_{p^*}}, ~ ~ \mu_{p^*} =  (\frac{1}{k'})^\frac{1}{2} \frac{1}{{\rm dn}^*  u_{p^*}} ,    
$$
with the Boltzmann weights:
$$ 
\begin{array}{lll}
W_{p^*q^*}(0)=1, & W_{p^*q^*}(1) = k'^{-1} {\rm scd}^* (u_{p^*}- u_{q^*} + K^*) &= {\rm exp}(-2J^*) ; \\
\overline{W}_{p^*q^*}(0)=1 , & \overline{W}_{p^*q^*}(1) = k'^{-1} {\rm scd}^* (u_{q^*}- u_{p^*} )&= {\rm exp}(-2\overline{J^*}),
\end{array}
$$
where the Jacobi theta functions are of modulus $\frac{{\rm i} k}{k'}$ with the elliptical integrals $(K^*, {K^*}') = (K (\frac{{\rm i}k}{k'}), K'(\frac{{\rm i}k}{k'}) ) = (K'(\frac{1}{k'}), K (\frac{1}{k'}))$. When $p^*$ is the dual rapidity of $p$, one finds $u_{p^*} = k' u_p$, (equivalently $K^* = k'K $), and the equivalence of (\req(WW*)) with the duality of weights in Ising model (\cite{KW}, \cite{Bax} (6.2.14a)):
$$
\tanh \overline{J} = {\rm e}^{-2J^*} , ~ ~ \tanh J = {\rm e}^{-2\overline{J^*}} . 
$$
The relation (\req(TT*l)) is identified with the usual duality in Ising model (see, e.g. \cite{Bax} Chapter 6). 
Note that the value $\tau = {\rm i} \frac{K}{K'}$ for modular $k'$ and the value $\tau^* = {\rm i} \frac{K^*}{{K^*}'}$ for $k'^{-1}$ are related by  $\tau^* = \frac{-\tau }{\tau + 1}$ when $k'$ are in a certain region, e.g. for real and positive $k'$.

\subsection{ Duality and Onsager algebra symmetry in homogeneous chiral Potts model \label{ssec.DuOA}}
In this subsection, we show the duality symmetry of homogeneous CPM compatible with the associated quantum spin chain Hamiltonian. In particular, we obtain the explicit relation between the duality and the Onsager-algebra symmetry in the superintegrable case. For a general homogeneous CPM (\req(homCP)), the quantum spin chain Hamiltonian is obtained from $\widehat{T}(q)$ by letting $q \rightarrow p$ in ${\goth W}_{k'}$ with small $u$ up to the first order: 
\bea(lll)
\frac{x_q}{x_p}= 1-2k' (\frac{y_p}{x_p \mu_p^2})^\frac{N}{2}u , & \frac{y_q}{y_p}=  1+ 2k' (\frac{x_p \mu_p^2}{y_p})^\frac{N}{2} u , & \frac{\mu_q}{\mu_p}= 1-2k (x_p y_p)^\frac{N}{2} u , 
\elea(uapr)
which in turn yields 
$$
\begin{array}{l}
\widehat{T}(q) = 1+ 2L u \sum_{j=1}^{N-1}\frac{(x_py_p^{-1})^{j-\frac{N}{2}}}{1-\omega^{-j}} + u {\cal H}(k' ;p) + O(u^2) , \\
{\cal H} (k';p) = - \sum_{\ell=1}^L \sum_{j=1}^{N-1}( \frac{{\rm e}^{{\rm i} \phi (2j-N)/N } }{\sin \pi j/ N } Z^j_\ell Z^{-j}_{\ell+1} + k' \frac{{\rm e}^{{\rm i} \overline{\phi} (2j-N)/N }}{\sin \pi j/ N } X^j_\ell ),
\end{array}
$$
where ${\rm e}^\frac{2{\rm i} \phi }{N}= (-1)^\frac{\pi {\rm i}}{N}\frac{x_p}{y_p}$ and ${\rm e}^\frac{2{\rm i} \overline{\phi} }{N}= (-1)^\frac{\pi {\rm i}}{N} \frac{x_p \mu_p^2}{y_p} $ (\cite{AMP} (1.10)-(1.17)). 
Note that in the superintegrable case (\req(pcood)), $u$ in (\req(uapr)) and the above ${\cal H} (k';p)$ are related to $\epsilon$ in (\req(xyep)) and $H(k')$ in (\req(H01)) by $(-1)^m u= \epsilon, ~ (-1)^m {\cal H} (k';p)=  H(k')$. Then the dual correspondence (\req(pp*)) between ${\goth W}_{k'}$ and ${\goth W}_{k'^{-1}}$ gives rise to the identification of $u$ and $\phi, \overline{\phi}$ at $p$ with those at the dual rapidity $p^*$:
$$
k' u =  u^* , ~ ~ {\rm e}^\frac{2{\rm i} \phi }{N} = {\rm e}^\frac{2{\rm i} \overline{\phi}^* }{N}, ~ ~ {\rm e}^\frac{2{\rm i} \overline{\phi} }{N} = {\rm e}^\frac{2{\rm i} \phi^* }{N} .
$$
Using (\req(uapr)) and the equality $\sum_{k=1}^{N-1} (N-k) \omega^{jk} = \frac{-N}{1-\omega^{-j}}$, one finds 
$$
W^{(f)}_{p q}(0) = \sqrt{N}(1 +2 \sum_{j =1}^{N-1} \frac{(x_py_p^{-1})^{j-\frac{N}{2} }}{1-\omega^{-j}}\epsilon ).
$$ 
By the second equality in (\req(TT*)), we obtain the explicit relation between Hamiltonian ${\cal H}(k'; p)$ and ${\cal H}(k'^{-1}; p^*)$: ${\cal H}(k'; p) = k' \Psi^{-1} {\cal H}(k'^{-1}; p^*) \Psi$. Indeed, by the relation between $\phi, \phi^*$ and  $\overline{\phi}^*, \phi^*$, one can verify this Hamiltonian relation directly by using (\req(Psi)) and the correspondence of local operators:
$$
Z_\ell Z_{\ell+1}^{-1} = \Psi^{-1} X^*_\ell \Psi , ~ ~ ~ X_{\ell+1} = \Psi^{-1} (Z^*_\ell Z_{\ell+1}^{* -1}) \Psi ~ ~ ~ ~  (1 \leq \ell \leq L).
$$

We now consider the homogeneous superintegrable case with the vertical rapidity $p$ in (\req(pcood)). As in section \ref{ssec.OACP}, the chiral-Potts transfer matrix $T_p(q)$ and $\widehat{T}_p(q)$ with $q \in {\goth W}_{k'}$  are expressed by (\req(TTform)) using the variables ${\tt x}, {\tt y}, \mu$ and quantum numbers $P_a, P_b, P_{\mu}$ . Under the duality transformation (\req(pp*)), $T_p, \widehat{T}_p$ are related to the transfer matrices $T_{p^*}(q^*), \widehat{T}_{p^*}(q^*)$ for $q^* \in {\goth W}_{k'^{-1}}$ with the vertical rapidity $p^*$ defined by
$$
p^*: (x_{p^*}, y_{p^*}, \mu_{p^*})= (\eta^{* \frac{1}{2}}\omega^{m+n_0},  \eta^{* \frac{1}{2}}\omega^{-n_0}, \omega^{n^*_0}) \in {\goth W}_{k'^{-1}},  
$$
where $\eta^* := (\frac{1-k^{' -1}}{1+k^{' -1}})^{\frac{1}{N}}$ and $n^*_0: = -n_0$. By Remark (1) at the end of section \ref{ssec.OACP}, $T_{p^*}(q^*), \widehat{T}_{p^*}(q^*)$ are again expressed by formulas in (\req(TTform)) with the variables ${\tt x}^*, {\tt y}^*, \mu^*$ and quantum numbers $P_a^*, P_b^*, P_{\mu^*}$, where the integer $m^*$ is defined by $m^* \equiv m+ 2n_0 \pmod{N}$ and $0 \leq m^* \leq N-1$. By Theorem \ref{thm:dualCP} and  the second relation in (\req(Pab)), $T_p(q), \widehat{T}_p(q)$ are equivalent to $T_{p^*}(q^*), \widehat{T}_{p^*}(q^*)$ via the linear isomorphism $\Psi$ (\req(Psi)) with $(Q, r) = (r^*, Q^*)$. Furthermore, the comparison of ${\tt x}, {\tt y}, \mu$ zero-orders for both sides of (\req(TT*)) (including those from $W^{(f)}_{p ,q}(0)^L, W^{(f)}_{p^* ,q^*}(0)^L $-factor) in turn yields the identification of variables and quantum numbers:
\bea(llll)
{\tt x}^* = \omega^{n_0} {\tt x} \mu, & {\tt y}^* = \omega^{n_0} {\tt y} \mu^{-1}, & \mu^* = \mu^{-1} , & {\tt t}^* = \omega^{2n_0} {\tt t}, \\
P_a^*= P_a , & P^*_b = P_b, & P_\mu \equiv r , & P_\mu^* \equiv Q , \\ 
 J^*= J, & m^*_E = m_E, &  \alpha^*_1 = \alpha_1 \omega^{n_0(P_b+P_a)},    
\elea(duqn)
as well as the identification of Bethe-polynomial, ${\tt F}({\tt t}) = {\tt F}^*({\tt t}^*)$, and their roots, ${\tt v}_j  = {\tt v}^*_j \omega^{2n_0} $, together with ${\cal G}^*(\lambda^*) = \lambda^{m_E} {\cal G} (\lambda)$ and $\overline{w}_i = - \overline{w^*}_i$. 
Note that by ${\tt F}(\omega^m ) = {\tt F}^*(\omega^{m^*} )$, the formula (\req(SR)) implies $S_R = S_R^*$, which is consistent with the $S_R$-equivariant-property of $\Psi$ in the remark of Proposition \ref{prop:t=t*}. Indeed, the isomorphism $\Psi$ makes the identification of basis elements in (\req(EF)), (\req(Ek')):
\be
\Psi: \vec{v}(s_1, \ldots, s_{m_E}; k') \mapsto \vec{v}(s_1, \ldots, s_{m_E}; k'^{-1}) . 
\ele(vec*)
The equivalence of ${\cal H}(k'; p)$ and ${\cal H}(k'^{-1}; p^*)$ now becomes the identification of Onsager-algebra generators:
\be
H_0 = \Psi^{-1} H^*_1 \Psi , ~ ~ H_1 = \Psi^{-1} H^*_0 \Psi ,
\ele(H0inf) 
where $(H_0, H_1), (H_0^*, H^*_1) $ are defined in (\req(H01)) for $(m, n_0), (m^*, n_0^*)$ and the boundary condition $r, r^*$ respectively. \par \vspace{1mm} \noindent
{\bf Remark.} In the special superintegrable case when $m=n_0=0$, the duality has been discussed by Baxter in \cite{B89}, where the variables $x, y, \mu$ in (2.1) and $x_d, y_d, \mu_d$ in (5.4) there correspond to ${\tt x}, {\tt y}^{-1}, {\tt y} \mu^{-1}$ and ${\tt x}^*, {\tt y}^{* -1}, {\tt y}^* \mu^{* -1}$ respectively in this paper.

\section{Inhomogeneous XXZ chain of  $U_q (sl_2)$-cyclic representation \label{sec:Uqsl}}
\setcounter{equation}{0}
For an arbitrary $q$, the quantum group $U_q (sl_2)$ is the $\CZ$-algebra generated by $K^\frac{\pm 1}{2}, e^{\pm}$ with $K^\frac{1}{2} K^\frac{-1}{2} = K^\frac{- 1}{2} K^\frac{1}{2} =1$  and the relation
\be 
 K^{\frac{1}{2}} e^{\pm } K^\frac{-1}{2}  =   q^{\pm 1} e^{\pm}  , ~ 
~ [e^+ , e^- ] = \frac{K-K^{-1}}{q - q^{-1}}. 
\ele(Uq)
Using $U_q (sl_2)$, one constructs a two-parameter family of  $L$-operators 
\be
{\bf L} (s)   =  \left( \begin{array}{cc}
         \rho^{-1} \nu^\frac{1}{2} s K^\frac{-1}{2}   -  \nu^\frac{-1}{2} s^{-1} K^\frac{1}{2}   &  (q- q^{-1}) e^-    \\
        (q - q^{-1}) e^+ &  \nu^\frac{1}{2} s K^\frac{1}{2} -  \rho \nu^\frac{-1}{2} s^{-1} K^\frac{-1}{2} 
\end{array} \right)
\ele(6vL)
with $\rho , \nu \neq 0 \in \CZ$, satisfying the YB equation 
\be
R_{\rm 6v} (s/s') ({\bf L}(s) \bigotimes_{aux}1) ( 1
\bigotimes_{aux} {\bf L}(s')) = (1
\bigotimes_{aux} {\bf L}(s'))( {\bf L}(s)
\bigotimes_{aux} 1) R_{\rm 6v} (s/s')
\ele(6YB)
for the symmetric six-vertex $R$-matrix \cite{Fad, KRS}:
$$
R_{\rm 6v} (s) = \left( \begin{array}{cccc}
        s^{-1} q - s q^{-1}  & 0 & 0 & 0 \\
        0 &s^{-1} - s & q - q^{-1} &  0 \\ 
        0 & q -q^{-1} &s^{-1} - s & 0 \\
     0 & 0 &0 & s^{-1} q - s q^{-1} 
\end{array} \right) .
$$   
For a chain of size $L$, we assign the $L$-operator ${\bf L}_\ell$ at the $\ell$th site with the parameter $\rho_\ell , \nu_\ell$ in (\req(6vL)). The monodromy matrix
\be
\bigotimes_{\ell=1}^L {\bf L}_\ell (s)  =  \left( \begin{array}{cc}
        {\bf A} (s)  & {\bf B} (s) \\
        {\bf C} (s) & {\bf D} (s)
\end{array} \right) 
\ele(M6V)
again satisfies YB (\req(6YB)), whose $q^{-2r}$-twisted trace 
\be
{\bf t} (s) = {\bf A} (s) + q^{-2r} {\bf D} (s),
\ele(t6v)
form the commuting family in the tensor algebra $( \stackrel{L}{\bigotimes} U_q (sl_2)) (s)$.
The leading and lowest terms of the monodromy matrix
$$
\begin{array}{ll}
{\sf A}_+ = \lim_{s  \rightarrow \infty} \widetilde{\nu}^\frac{-1}{2} \widetilde{\rho} s^{-L}  {\bf A}(s) , & 
{\sf A}_- = \lim_{s \rightarrow 0} \widetilde{\nu}^\frac{1}{2}(-s)^L  {\bf A}(s) ,   \\ 
{\sf B}_\pm = \lim_{s^{\pm 1} \rightarrow \infty}  \widetilde{\nu}^{\mp \frac{ 1}{2}} (\pm s)^{\mp (L-1)} \frac{{\bf B}(s)}{ q-q^{-1}} ,  & 
{\sf C}_\pm = \lim_{s^{\pm 1} \rightarrow \infty} \widetilde{\nu}^{\mp \frac{ 1}{2}} (\pm s )^{\mp (L-1)}\frac{{\bf C}(s)}{q-q^{-1}},  \\
 {\sf D}_+ = \lim_{s  \rightarrow \infty} \widetilde{\nu}^\frac{-1}{2} s^{- L}{\bf D}(s), & {\sf D}_- = \lim_{s \rightarrow 0} \widetilde{\nu}^\frac{1}{2} \widetilde{\rho}^{-1} (- s)^L {\bf D}(s), 
\end{array}  
$$ 
give rise to the quantum affine algebra $U_q (\widehat{sl}_2)$, where $
\widetilde{\nu} := \prod_\ell \nu_\ell$ and $\widetilde{\rho} := \prod_\ell \rho_\ell$. 
Indeed the generators of $U_q(\widehat{sl}_2)$,
$$
k_0^{-1} =k_1  = {\sf A}^2_- = {\sf D}^2_+  , ~ \ e_1  = S^+ (:= {\sf C}_+), \  f_1  = S^- (: = {\sf B}_-),  ~ \ e_0 = T^- (:= {\sf B}_+) , ~  f_0 = T^+ (:= {\sf C}_-), 
$$
are expressed by
\bea(l)
{\sf A}_- = {\sf D}_+ = K^{\frac{1}{2}} \otimes \cdots \otimes K^{\frac{1}{2}}, \ ~ ~ \ ~
{\sf A}_+ = {\sf D}_- = K^{ \frac{-1}{2}} \otimes \cdots \otimes K^{\frac{-1}{2}}, \\
S^\pm = \sum_{i=1}^L  \nu_i^{\mp \frac{1}{2}} (\prod_{j >i} \rho_j)^{\mp 1}  \underbrace{K^{\frac{1}{2}} \otimes \cdots \otimes K^{\frac{1}{2}}}_{i-1}\otimes e^\pm \otimes  \underbrace{K^{ \frac{-1}{2}} \otimes \cdots \otimes K^{ \frac{-1}{2}}}_{L-i} , 
\\
T^\pm  =  \sum_{i=1}^L  \nu_i^{\pm \frac{1}{2}} (\prod_{j < i} \rho_j)^{\pm 1}  \underbrace{K^{\frac{-1}{2}} \otimes \cdots \otimes K^{ \frac{-1}{2}}}_{i-1}\otimes e^\pm \otimes  \underbrace{K^{\frac{1}{2}} \otimes \cdots \otimes K^{\frac{1}{2}}}_{L-i} . 
\elea(STpm)
In the homogeneous case with the periodic boundary condition, i.e., $r=0, \nu = \nu_\ell, \rho = \rho_\ell$ for all $\ell$,  the quantum group $U_q(\widehat{sl}_2)$ also possesses a Hopf-algebra structure defined by
$$
\begin{array}{ll}
\bigtriangleup (k_i) = k_i \otimes k_i, ~ ~ ~ i=0, 1 ,& \\
\bigtriangleup (e_1 ) =  k^\frac{1}{2}_1 \otimes e_1    +   \rho^{- 1} e_1 \otimes k^\frac{1}{2}_0 , &
\bigtriangleup (f_1 ) =  k^\frac{1}{2}_1 \otimes f_1    +  \rho f_1 \otimes k^\frac{1}{2}_0 , \\
\bigtriangleup (e_0 ) =  \rho^{-1} k^\frac{1}{2}_0 \otimes e_0    +  e_0 \otimes k^\frac{1}{2}_1 , &
\bigtriangleup (f_0 ) =  \rho k^\frac{1}{2}_0 \otimes f_0    +  f_0 \otimes k^\frac{1}{2}_1 .
\end{array}
$$
In particular, with $\rho =1 , \nu = q^{d-2}$ and the spin-$\frac{d-1}{2}$ (highest weight) representation of $U_q (sl_2)$ on  $\CZ^d = \oplus_{k=0}^{d-1} \CZ {\bf e}^k $:
\be
K^{\frac{1}{2}} ({\bf e}^k) = q^{\frac{d-1-2k}{2}} {\bf e}^k , \ \ e^+ ( {\bf e}^k ) = [ k  ] {\bf e}^{k-1} , \ \ e^-( {\bf e}^k ) = [ d-1-k ] {\bf e}^{k+1},
\ele(spinrp)
where $[n]= \frac{q^n - q^{-n}}{q- q^{-1}}$ and $ e^+ ( {\bf e}^{0} ) = e^- ( {\bf e}^{d-1} )= 0$, (\req(t6v)) gives rise to the transfer matrix of the well-known homogeneous XXZ chain of spin-$\frac{d-1}{2}$ (see, e.g. \cite{KiR, R06Q, R06F} and references therein).

Hereafter in this paper, we assume the anisotropic parameter $q$ of the inhomogeneous model (\req(M6V)) to be a $N$th primitive root of unity. There is a three-parameter family of $U_q ( sl_2)$-cyclic representation on $\CZ^N$ (the space of cyclic $N$-vectors), $s_{\phi, \phi^\prime, \varepsilon}$, labeled by non-zero complex numbers $\phi, \phi^\prime$ and $\varepsilon$, and defined  by\footnote{The definition (\req(crep)) in this paper is the same as \cite{R0806} (3.1) where $|n \rangle$  is replaced by $\widehat{|-\sigma} \rangle$ here.}
\bea(ll)
K^\frac{1}{2} \widehat{|\sigma } \rangle =  q^{\sigma+\frac{\phi^\prime-\phi}{2}} \widehat{|\sigma} \rangle , & \\
e^+  \widehat{|\sigma} \rangle =  q^{ \varepsilon} \frac{  q^{\phi - \sigma}-  q^{- \phi + \sigma}  }{q-q^{-1}} \widehat{|\sigma + 1 }\rangle, &
e^-  \widehat{|\sigma} \rangle =  q^{ -\varepsilon} \frac{ q^{\phi^\prime  +\sigma}-  q^{- \phi^\prime  - \sigma}  }{q-q^{-1}} \widehat{|\sigma - 1} \rangle ,  \\
\elea(crep)
(see, e.g. \cite{DJMM, DK}), where $\widehat{|\sigma} \rangle ~ (\sigma \in \ZZ_N)$ are the Fourier basis of $\CZ^N$ in (\req(Fb)). 
Applying the cyclic representation $s_{\phi_\ell, \phi_\ell^\prime, \varepsilon_\ell}$ on ${\bf L}_\ell$ in (\req(M6V)), we form the monodromy matrix 
\be
\bigotimes_{\ell=1}^L {\cal L}_\ell (s)  =  \left( \begin{array}{cc}
        {\cal A} (s)  & {\cal B} (s) \\
        {\cal C} (s) & {\cal D} (s)
\end{array} \right),  ~ ~ {\cal L}_\ell (s) = s_{\phi_\ell , \phi_\ell^\prime, \varepsilon_\ell} {\bf L}_\ell(s).
\ele(6vM)
The transfer matrices of the inhomogeneous XXZ chain of $U_q (sl_2)$-cyclic representation with parameters $\{\phi_\ell, \phi_\ell^\prime, \varepsilon_\ell, \nu_\ell, \rho_\ell \}_\ell$ and with the boundary condition (\req(sBy))
are the $\stackrel{L}{\otimes}\CZ^N$-operator
\be
T (s) =   {\cal A} (s) + q^{-2r} {\cal D} (s) =  (\otimes_\ell s_{\phi_\ell , \phi_\ell^\prime, \varepsilon_\ell}) {\bf t}(s),
\ele(TcXZ)
which commute with $K^\frac{1}{2} (:= \otimes_\ell K^\frac{1}{2}_\ell$, the product of local $K^\frac{1}{2}$-operators ). 

In the case when $\phi_\ell, \phi'_\ell $'s are integers for all $\ell$, we write the basis elements in (\req(crep)) by 
\be
v^k (= v^k_\ell) := \widehat{| \phi_\ell + N - k} \rangle , ~ ~ \overline{v}^k (= \overline{v}^k_\ell) := \widehat{|-\phi'_\ell -1 - k} \rangle ~ ~ ~ (0 \leq k \leq N-1). 
\ele(vwk)
Then $K (v^k) = q^{\phi_\ell + \phi'_\ell - 2k} v^k$ and $K ( \overline{v}^k ) = q^{-\phi_\ell - \phi'_\ell - 2k-2} \overline{v}^k$. With $K = q^{h_\ell}$, the relation (\req(crep)) is consistent with
$$
\begin{array}{ll}
e^+ (v^k) = q^{\varepsilon_\ell } [k] v^{k-1}, & [h_\ell, e^+] (v^k) = (h_\ell (v^{k-1}) - h_\ell (v^k) ) e^+ (v^k), \\
e^- (\overline{v}^k) = q^{-\varepsilon_\ell } [N-k-1] \overline{v}^{k+1}, & [h_\ell, e^-](\overline{v}^k) = (h_\ell (\overline{v}^{k+1}) - h_\ell (\overline{v}^k) ) e^+ (\overline{v}^k),
\end{array}
$$
which in turn yield $h_\ell (v^{k-1}) = h_\ell (v^k)+ 2$ and $h_\ell (\overline{v}^{k-1})- 2 = h_\ell (\overline{v}^{k}) $ for $1 \leq k \leq N-1$, (note that $e^+ (v^0) = e^- (\overline{v}^{N-1}) = 0$). Therefore $v^k = \overline{v}^k$, equivalently   
\be
\phi_\ell , ~ \phi'_\ell \in \ZZ , ~ ~ ~ ~ \phi_\ell + \phi'_\ell + 1 \equiv 0 \pmod{N} , 
\ele(phih)
hence $h_\ell (v^k) = -1 - 2k + c_\ell N ~ (1 \leq k \leq N-1)$ for an integer $c_\ell$. When $\phi_\ell , \phi'_\ell$ satisfy the condition (\req(phih)), one may define the normalized $N$th power of $S^\pm , T^\pm$, $S^{\pm (N)}= \frac{S^{\pm N}}{[N]!}$, $T^{\pm (N)}= \frac{T^{\pm N}}{[N]!}$, with the expression:
\bea(ll)
S^{\pm (N)} =& \sum_{ 0 \leq k_i < N, \  k_1+\cdots+ k_L=N } \frac{1}{[k_1]! \cdots [k_L]!} \otimes_{i=1}^L  K_i^{\frac{-1}{2}( \sum_{j (<i)} - \sum_{j (>i)})k_j  } e_i^{\pm  k_i} \nu_i^{ \frac{\mp k_i}{2}} \rho_i^{\mp \sum_{j (<i)} k_j} , \\
T^{\pm (N)} =& \sum_{ 0 \leq k_i < N, \  k_1+\cdots+ k_L=N } \frac{1}{[k_1]! \cdots [k_L]!} \otimes_{i=1}^L  K_i^{\frac{1}{2}(\sum_{j (<i)} - \sum_{j (>i)})k_j  }  e_i^{\pm k_i} \nu_i^{ \frac{\pm k_i}{2}} \rho_i^{\pm \sum_{j (>i)} k_j} .
\elea(STNpm)
As in \cite{R06F} section 4.2, the entries of (\req(6vM)) satisfy the ABCD algebra, which in turn yields 
$$
\begin{array}{c} 
{\cal A}(s) \prod_{i=1}^N {\cal B}(s_i) =  \frac{ x^2  -s^2 q^{N+1}}{x^2 q^{N} - s^2 q }  \prod_{i=1}^N {\cal B}(s_i) {\cal A}(s) +  \frac{s s_1 (q^2 - 1)}{q (s_1 ^2- s^2 )} \frac{[N]!}{[N-1]!}  {\cal B} (s)  \prod_{i=2 }^N {\cal B}(s_i) {\cal A}(s_1), \\
{\cal D}(s) \prod_{i=1}^N {\cal B}(s_i)=  \frac{ x^2 -s^2 q^{-N-1} }{x^2 q^{-N} - s^2 q^{-1} }  \prod_{i=1}^N {\cal B}(s_i) {\cal D}(s) -  \frac{s s_N (q^2 -1)}{q (s_N^2- s^2 )}  \frac{[N]!}{[N-1]!} {\cal B}(s) \prod_{i=1}^{N-1} {\cal B}(s_i) {\cal D}(s_N)  , 
\end{array}
$$
as well as another two formulas by interchanging the above ${\cal A, B}$ and ${\cal D, C}$ respectively, where $ s_i = x q^{\frac{N+1-2i}{2}}$ for $i = 1 , \ldots, N$. By multiplying  $(\pm s_i)^{\mp (L-1)} \widetilde{\nu}^{\mp \frac{ 1}{2}} (q-q^{-1})^{-1}$ to ${\cal B}(s_i)$ or ${\cal C}(s_i)$ in above at a generic $q$, then taking the limit of $s_i^{\pm 1}$ at $q^N=1$, one arrives
$$
\begin{array}{lll} 
&[{\cal A}(s),  S^{\pm (N)}]=  \mp  s^{\pm 1} S^{\pm (N-1)} {}^{{\cal C}(s)\widetilde{\rho}^{-1}}_{{\cal B}(s)}   K^{ \frac{\mp 1}{2}},  & [{\cal D}(s),  S^{\pm (N)}]=  \pm  s^{\pm 1} S^{\pm (N-1)} {}^{{\cal C}(s)}_{{\cal B}(s)\widetilde{\rho}}   K^{ \frac{\pm 1}{2}}  ,  \\
&[{\cal A}(s),  T^{\pm (N)}]=  \mp  s^{\mp 1} T^{\pm (N-1)} {}^{{\cal C}(s)}_{{\cal B}(s)\widetilde{\rho}^{-1}}   K^{ \frac{\pm 1}{2}},  & [{\cal D}(s),  T^{\pm (N)}]=  \pm  s^{\mp 1} T^{\pm (N-1)} {}^{{\cal C}(s)\widetilde{\rho}}_{{\cal B}(s)}   K^{ \frac{\mp 1}{2}}  ,  \\
\end{array}
$$
which imply 
$$
\begin{array}{ll}
& [T (s), S^{\pm (N)}  ]= s^{\pm 1}  S^{\pm (N-1)} {}^{{\cal C}(s)}_{{\cal B}(s)}   ( {}^{q^{-2r}}_{1}  K^{ \frac{1}{2}} - {}^{\widetilde{\rho}^{-1}}_{q^{-2r} \widetilde{\rho}}   K^{ \frac{-1}{2}}), \\
&   [ T(s), T^{\pm  (N)}  ] = -s^{\mp 1} T^{\pm  (N-1)} {}^{{\cal C}(s)}_{{\cal B}(s)}  ( {}^{1}_{q^{-2r}}  K^{ \frac{1}{2}} - {}^{q^{-2r} \widetilde{\rho}}_{\widetilde{\rho}^{-1}}   K^{ \frac{-1}{2}}).
\end{array}
$$
In particular, under the constraint $K = q^{-2r} \widetilde{\rho}$ with $q^{-2r} \widetilde{\rho} = \pm 1$ ,  $T^{\pm (N)}$ and $S^{\pm (N)}$ commute with $T(s)$ in (\req(TcXZ)):
\be
[T (s), S^{\pm (N)}  ] = [T (s), T^{\pm (N)}  ] = 0 , \ \ {\rm when} ~  K = q^{-2r} \widetilde{\rho}= \pm 1 .
\ele(tSTcom)
In this situation, with $H^{(N)}= N^{-1} \sum_{\ell=1}^L 1 \otimes \cdots \otimes h_\ell \otimes 1 \otimes \cdots \otimes 1$, the operators, $H^{(N)}, S^{\pm (N)}$ and $T^{\pm (N)}$,  generate a $sl_2$-loop-algebra representation,
$$
~ [ H^{(N)} , S^{\pm (N)} ] = \pm 2  S^{\pm (N)}, ~ ~ [ H^{(N)} , T^{\pm (N)} ]= \pm 2 T^{\pm (N)} 
$$
with the Chevalley generators, 
\bea(l)
- H_0 = H_1 = H^{(N)}, E_1= S^{+ (N)}, F_1 = S^{- (N)}, E_0 = T^{- (N)},  F_0 = T^{+ (N)},
\elea(loopb)
which are related to the mode basis by $E_1= e(0), F_1 =f(0), E_0 =f(1),  F_0 =e(-1)$.
Note that the representation $s_{\phi_\ell, \phi'_\ell, \varepsilon_\ell }$ in (\req(phih)) is  equivalent to the spin-$\frac{N-1}{2}$ representation in (\req(spinrp)) with the basis $v^k$'s  corresponding to ${\bf e}^k$'s. In this situation, the Algebraic-Bethe-Ansatz method \cite{Fad, KBI, KS} can be employed to the diagonalization of the transfer matrix (\req(TcXZ)).

\subsection{The equivalence of inhomogeneous $\tau^{(2)}$-models and XXZ chains with $U_q (sl_2)$-cyclic representation \label{ssec.equiv}}
As in the homogeneous case \cite{R0806}, we now show the general inhomogeneous $\tau^{(2)}$-models  (\req(tau2)) are equivalent to XXZ chains in (\req(TcXZ)). For simplicity, we consider only the odd $N (=2M+1)$ case\footnote{For an arbitrary $N$, the argument in the homogenous case  \cite{R0806} section 3.2 can also be applied to the general inhomogeneous case so that a XXZ chain (\req(TcXZ)) is equivalent to the sum of two copies of the same $\tau^{(2)}$-model (\req(tau2)) via the identification of parameters (\req(Par=1)).} where $q$ is chosen to be the $N$th root-of-unity satisfying $q^{-2} = \omega$, i.e, $q:= \omega^M$. One can express the cyclic representation (\req(crep)) in terms of $\widehat{X}, \widehat{Z}$: 
$$
\begin{array}{lll}
K^\frac{1}{2} = q^\frac{\phi^\prime- \phi}{2}  \widehat{Z}^\frac{-1}{2},& 
e^+  = q^{ \varepsilon} \frac{(  q^{\phi +1} \widehat{Z}^{\frac{ 1}{2}} -  q^{- \phi -1} \widehat{Z}^{  \frac{- 1}{2}})\widehat{X}}{q-q^{-1}} &
e^-  =q^{ -\varepsilon} \frac{(  q^{\phi^\prime +1} \widehat{Z}^{  \frac{- 1}{2}} -  q^{- \phi^\prime -1} \widehat{Z}^{ \frac{ 1}{2}})\widehat{X}^{-1}}{q-q^{-1}}.
\end{array}
$$
By (\req(XZF)), the operators $K^{-1}, K^\frac{-1}{2} e^\pm$ are represented in the form    
\bea(ll)
K^{-1} = q^{\phi - \phi^\prime} X , &  \\
K^\frac{-1}{2} e^+  = -q^{\frac{-\phi - \phi^\prime}{2} + \varepsilon-1} \frac{( 1 -q^{2\phi+2}X   )Z^{- 1}}{q-q^{-1}}, &
K^\frac{-1}{2} e^-  = q^{\frac{\phi+\phi^\prime}{2} - \varepsilon+1} \frac{(  1   -  q^{-2\phi^\prime -2}X )Z}{q-q^{-1}}. 
\elea(crXZ) 
By setting $t = s^2$ and with the gauge transform ${\rm dia}[1, -sq]$, the modified $L$-operator, $-s \nu_\ell^\frac{1}{2} K_\ell^{\frac{-1}{2}} {\cal L}_\ell (s)$, of (\req(6vM)) is equivalent to 
$$
\left( \begin{array}{cc}
        1- t \rho_\ell^{-1} \nu_\ell q^{\phi_\ell - \phi_\ell^\prime}       X   &  \nu_\ell^\frac{1}{2} q^{\frac{\phi_\ell+\phi_\ell^\prime}{2}- \varepsilon_\ell} ( 1  -  q^{- 2 \phi_\ell^\prime -2} X ) Z   \\
         - t \nu_\ell^\frac{1}{2} q^{\frac{-\phi_\ell - \phi_\ell^\prime}{2}+ \varepsilon_\ell} (  1-  q^{2 \phi_\ell +2} X ) Z^{- 1} & - t \nu_\ell   +    \rho_\ell q^{\phi_\ell - \phi_\ell^\prime}    X
\end{array} \right) ,
$$
which is the same as $L_\ell (t)$ in (\req(tau2)) with the identification of parameters:
\be
{\sf a}_\ell = \rho_\ell \nu_\ell^\frac{-1}{2}  q^{\frac{-\phi_\ell - \phi_\ell^\prime}{2} - \varepsilon_\ell} , ~ \omega {\sf a}_\ell {\sf a'}_\ell = \rho_\ell^2 \nu_\ell^{-1}, ~  {\sf b}_\ell  = \nu_\ell^\frac{-1}{2} q^{\frac{-\phi_\ell - \phi_\ell^\prime}{2}+ \varepsilon_\ell},  ~   {\sf b }_\ell{\sf b'}_\ell = \nu_\ell^{-1},  ~ {\sf c}_\ell  = \rho_\ell^{-1} q^{\phi_\ell - \phi_\ell^\prime} ,  
\ele(Par=1)
equivalently, $
q^{\varepsilon_\ell} = (\frac{\omega {\sf a}'_\ell{\sf b}_\ell}{{\sf a}_\ell {\sf b}'_\ell })^\frac{1}{4}$, $q^{2 \phi_\ell }= \frac{\omega {\sf a}_\ell' {\sf c}_\ell}{{\sf b}_\ell}$ , $q^{2 \phi_\ell'}= \frac{{\sf b}_\ell'}{{\sf a}_\ell {\sf c}_\ell}$ , $\rho_\ell^2 = \frac{\omega {\sf a}_\ell {\sf a}'_\ell}{{\sf b}_\ell {\sf b}'_\ell}$, $\nu_\ell=  \frac{1}{{\sf b}_\ell {\sf b}'_\ell}$. 
Therefore $\tau^{(2)}$-models (\req(tau2)) are equivalent to XXZ chains (\req(TcXZ))  in the general inhomogeneous case, where the product of local operator $K^{-1}$'s corresponds to (a scalar multiple of) the spin-shift operator $X$ in $\tau^{(2)}$-model. Note that with ${\sf a}',  {\sf b}', {\sf a}, {\sf b}, {\sf c}$ in (\req(pp'l)), the formula (\req(Par=1)) establishes a relation between the chiral Potts rapidities and cyclic-representation parameters, which gives a scheme of reproducing the Boltzmann weights of CPM from the representation theory of $U_q(sl_2)$ in \cite{DJMMb}.

\subsection{The connection between Onsager-algebra symmetry of superintegrable $\tau^{(2)}$-model and the $sl_2$-loop-algebra of XXZ chain \label{ssec.symEq}}
In this section, we consider a special homogeneous XXZ chain case in (\req(phih)), $
\nu = \nu_\ell ,  \rho = \rho_\ell ,  \varepsilon = \varepsilon_\ell ,  \phi= \phi_\ell ,  \phi'= \phi'_\ell$ for all $\ell $ with  integers $\phi, \phi'$ satisfying $\phi + \phi' + 1 \equiv 0 \pmod{N}$ for odd $N=2M+1$. As before, $q^{-2}= \omega$ with $q= \omega^\frac{-1}{2} (:= \omega^M)$.  By (\req(Par=1)) and identifying the spectral parameter ${\tt t}= s^2$ , the superintegrable $\tau^{(2)}$-model (\req(hsupL))  is equivalent to the homogeneous XXZ chain with $\rho = \omega^{m-M}$, $\nu= 1$,  $\phi \equiv -1-m-2n_0$, $ \phi^\prime \equiv m+2n_0$, $\varepsilon= M$ in (\req(6vM)):
\be 
{\cal L}_\ell (s)   =  \left( \begin{array}{cc}
         q^{1+2m} s K^\frac{-1}{2}   -   s^{-1} K^\frac{1}{2}   &  (q- q^{-1}) e^-    \\
        (q - q^{-1}) e^+ &   s K^\frac{1}{2} -  q^{-1-2m}   s^{-1} K^\frac{-1}{2} 
\end{array} \right) 
\ele(xxzsp) 
for all $\ell$. Here  $K, e^\pm$ are in (\req(crXZ)), expressed by 
$$
\begin{array}{ll}
K^\frac{1}{2} =  q^{\frac{1}{2} + m+2n_0}\widehat{Z}^\frac{-1}{2}, & \\
e^+  = q^\frac{-1}{2} \frac{(q^{-m-2n_0}\widehat{Z}^\frac{1}{2}   -q^{m+2n_0 } \widehat{Z}^\frac{-1}{2}  )}{q-q^{-1}}\widehat{X} ,
& e^- =q^\frac{1}{2} \frac{(q^{1+m+2n_0} \widehat{Z}^\frac{-1}{2}-  q^{-1-m-2n_0}\widehat{Z}^\frac{1}{2}) }{q-q^{-1}}\widehat{X}^{-1}, 
\end{array} 
$$
where $\widehat{X}, \widehat{Z}$ are the local Weyl operators (\req(XZF)). Write the second Onsager-algebra operator $H_1$ of (\req(H01)) in the form
$$
H_1 = -2 \sum_\ell S^z_\ell , ~ ~ ~ 
S^z (=S^z_\ell) := \sum_{j=1}^{N-1} \frac{ \omega^{(m + 2n_0 )j }\widehat{Z}_\ell^j }{1-\omega^{-j}} .
$$
By the equality, $\sum_{j=1}^{N-1} \omega^{kj}(1-\omega^{-j})^{-1} = (N-1 -2k)/2$  for $ 0\leq k \leq N-1$, one finds
\bea(ll)
S^z ({\bf e}^k) = \frac{N-1}{2} - k , ~ ~ ~ {\bf e}^k:= \widehat{|-m-2n_0+ k} \rangle ~ ~  ~ (k=0, \ldots, N-1).
\elea(ek)
Using the basis ${\bf e}^k$'s, the operators $K, e^\pm$ in  (\req(xxzsp)) are now expressed by
\bea(lll)
K^\frac{1}{2}({\bf e}^k) =  q^{-S^z} {\bf e}^k, &
e^+  ({\bf e}^k) = q^\frac{-1}{2} [N-1-k] {\bf e}^{k+1} , &
e^-({\bf e}^k)  =q^\frac{1}{2} [k] {\bf e}^{k-1},  
\elea(spN-1)
which is similar to the spin-$\frac{N-1}{2}$ (highest weight) $U_q(sl_2)$-representation (\req(spinrp)) via the linear isomorphism of $\CZ^N$, ${\bf e}^k \mapsto q^\frac{k}{2} {\bf e}^{N-1-k}$. Hence we have shown the following equivalent relation: 
\begin{lem}\label{lem:supspin}
The superintegrable $\tau^{(2)}$-model $(\req(hsupL))$  is equivalent to the homogeneous XXZ chain $(\req(xxzsp))$ with the spin-$\frac{N-1}{2}$~$U_q(sl_2)$-representation $(\req(spN-1))$. Furthermore the Onsager-algebra operator $\frac{-H_1}{2}$ in $(\req(H01))$ is the total $S^z$-operator with $-2S^z \equiv  (1+2m+4n_0)L+2Q$.
\end{lem}
In particular, when $m=M$  and $n_0=0$, the periodic XXZ chain (\req(xxzsp)) is equivalent to the usual spin-$\frac{N-1}{2}$ chain for the representation $(\req(spinrp))_{d=N}$, i.e. the homogeneous XXZ chain, (\req(6vL)) and (\req(crep)), with $ \phi = \phi^\prime =M, \varepsilon=0, \nu= \rho=1$ (\cite{R075} (4.5)-(4.9)), where the $sl_2$-loop-algebra symmetry is known to exist in the sector $2S^z \equiv 0 \pmod{N}$ \cite{NiD, R06F}. In the general case, by (\req(tSTcom)), the $\tau^{(2)}$-model $(\req(hsupL))$ possesses the $sl_2$-loop-algebra symmetry in the sector $K = q^{-2r} \rho^L = \pm 1$,  which by (\req(Pab)), is equivalent to  
\be
2S^z \equiv 2r +(2m+1)L \equiv 0  \Leftrightarrow  P_b = P_a \equiv Q + 2n_0 L - r  \equiv 0  \pmod{N}.
\ele(Sect)

We now employ the Algebraic-Bethe-Ansatz techniques to determine the Bethe states  as in the $m=M, n_0=r=0$ case (\cite{R06F} section 3). The pseudo-vacuum of the $\tau^{(2)}$-model (\req(hsupL)) are defined by 
\bea(lll)
\Omega^+ ~ (=\Omega^+_L) :=  \stackrel{L}{\otimes} {\bf e}^{N-1} & {\rm or} & \Omega^- ~ (=\Omega^-_L) :=  \stackrel{L}{\otimes} {\bf e}^0 , 
\elea(vac)
where ${\bf e}^k$'s are the basis in (\req(ek)). Then 
$$
\begin{array}{lll}
C({\tt t}) \Omega^+ = 0, &A({\tt t}) \Omega^+  = {\tt h}(\omega^{-1} {\tt t} ) \Omega^+, & D({\tt t}) \Omega^+ = \omega^{mL} {\tt h}({\tt t} ) \Omega^+ ; \\
B({\tt t}) \Omega^- = 0, & A({\tt t}) \Omega^- = {\tt h}({\tt t} ) \Omega^-, & D({\tt t})\Omega^-  = \omega^{(1+m)L} {\tt h}(\omega^{-1}{\tt t} ) \Omega^-,
\end{array}
$$
where $A, B, C, D$ are the entries of monodromy matrix (\req(Mont2)), and ${\tt h}({\tt t} )$ is the polynomial in (\req(hFv)). One may study the eigenvalues and eigenstates of the $\tau^{(2)}$-model using its ABCD-algebra structure (\cite{R06F} section 2). The Bethe states, $\psi^\pm (= \psi^\pm ({\tt v}^\pm_1, \ldots,{\tt v}^\pm_J))$,  are defined by\footnote{the results in the $m=n_0=r=0$ case were given in \cite{R06F}  section 3.1, where the vector $f_k$ is $\sqrt{N} \widehat{|-k} \rangle$ in this paper, and  $m, h_1(t)^L, h_2(t)^L, F(t), -t_j^{-1} $ in formulas (3.4)-(3.9) there corresponds respectively to $J, {\tt h}({\tt t} ), {\tt h}({\tt t} ), {\tt F}({\tt t}), \omega {\tt v}_j$ in this paper.}
\bea(ll)
\psi^+ = \prod_{j=1}^J B(-(\omega {\tt v}^+_j)^{-1}) \Omega^+, & (P^+_a=0, P^+_b \equiv mL +r -J ); \\
\psi^- = \prod_{j=1}^J C(-(\omega {\tt v}^-_j)^{-1}) \Omega^- , & (P^-_a\equiv -(1+m)L-r-J, P^-_b=0 ),  
\elea(Betv)
where ${\tt v}^\pm_j$'s form a solution of Bethe equation  (\req(Bethesup)) with the $\tau^{(2)}$-eigenvalue (\req(stauev)). Note that the $\ZZ_N$-charges of the above $\psi^+, \psi^- $ are $Q \equiv -(1+m+2n_0)L-J, - (m+2n_0)L+J$ respectively. We now identify the Bethe states in (\req(Betv)) with the basis elements at $k'= \infty$ in (\req(Ek')). By using (\req(hWw)), one finds the expression of $H_1$-eigenvalues from the formula (\req(Esk')) at $k'=\infty$, 
\bea(l)
-2S^z =  (N-1- 2m)L + 2(P_b - P_a - P_\mu)  + N \sum_{i=1}^{m_E}(s_i-1) .
\elea(Sz)
The pseudo-vacuum $\Omega^+$ (or $\Omega^-$) is a state in $\stackrel{L}{\otimes} \CZ^N$ with the maximum  (minimum respectively) $-2S^z$. 
As in the case of six vertex model at roots of unity \cite{De05, DFM, FM04}, the Bethe state $\psi^\pm $ in (\req(Betv)) is characterized as the vector with maximum or minimum $-2S^z$ in ${\cal E}_{{\tt F}^\pm, P^\pm_a , P^\pm_b }$, by which the relation (\req(Sz)) in turn yields $\psi^+ = \vec{v} (+, \ldots, +; \infty )$ or $\psi^- = \vec{v} (-, \ldots, -; \infty ) $.
The condition (\req(Sect)) is characterized as the sector with $P_a= P_a^\pm =0, P_b= P_b^\pm =0$ in (\req(Betv)), where we define the Bethe states $\psi^\pm$ with ${\tt v}_j = {\tt v}^\pm_j$ for all $j$. By a similar argument in the case of spin-$\frac{N-1}{2}$ XXZ chain (see, \cite{NiD} or \cite{R06F} section 4.2)\footnote{The variable $t(= qs^2)$ in \cite{R06F} Theorem 4 differs from $t$ in this paper by the factor $q$, but with the same $t^N$-polynomial: ${\tt P}({\tt t)}$ in this paper $= P_{6V}(t^N)$ in \cite{R06F} (4.32).}, the operators $S^{\pm (N)}, T^{\pm (N)}$ in (\req(STNpm)) and $H^{(N)} = -2S^z/N$ on ${\cal E}_{{\tt F}, P_a, P_b}$ form a $sl_2[z,z^{-1}]$-representation, having the evaluation polynomial ${\tt P}({\tt t})/N$ for the ${\tt t}^N$-polynomial ${\tt P}({\tt t)}$ in (\req(sPt)). 
The $sl_2$-loop-algebra generators are described by (\req(loopb)) when using $\psi^+$, and with the indices $0, 1$ interchanged in the case $\psi^-$. This $sl_2[z,z^{-1}]$-representation is induced from a $(\stackrel{m_E}{\oplus} sl_2)$-structure on ${\cal E}_{{\tt F}, P_a, P_b}$, then evaluating $z$ on roots ${\tt t}^N_i$ of ${\tt P}({\tt t)}$, together with the spin-$\frac{1}{2}$-representation of $sl_2$.  Note that as the $sl_2[z,z^{-1}]$-structures of ${\cal E}_{{\tt F}, P_a, P_b}$, the $sl_2$-loop-algebra symmetry induced from XXZ chain is different from $\rho_{\infty}$ in (\req(rhok')) where the evaluation-values ${\rm e}^{{\rm i} \theta_i} (\neq {\tt t}^N_i)$ are defined in (\req(OAev)). However, both $sl_2[z,z^{-1}]$-representations share the same highest or lowest weight vector, $\psi^\pm= \vec{v} (\ldots, \pm, \ldots ; \infty )$, hence they both give rise to the same underlying $(\stackrel{m_E}{\oplus} sl_2)$-structure of ${\cal E}_{{\tt F}, P_a, P_b}$, determined by $e^\pm z^k ~ ( 0 \leq k < m_E)$ on the highest or lowest weight vector, and presented by the basis in (\req(Ek')) at $k'=\infty$.

Using (\req(tf=t)) and the duality (\req(TT*)), we may also connect the Onsager-algebra symmetry at $k'=0$ with the $sl_2$-loop-algebra symmetry of XXZ chain through the $\tau^{(2)}$-face model. 
The $L$-operator (\req(FaLpp')) of a homogeneous superintegrable $\tau^{(2)}$-face model with a vertical rapidity $p$ (\req(pcood)) is expressed by ${\tt C}^N (= \sum_{n \in \ZZ_N} \CZ |n \rangle \rangle)$-Weyl-operators:
\bea(l)
{\tt L}_\ell ( t ) =   \left( \begin{array}{cc}
        1  -  {\tt t}  {\tt Z}   & (1  -\omega^{1+m} {\tt Z}) {\tt X}^{-1} \\
       -  \omega^{2n_0} {\tt t}(1 -  \omega^{m}{\tt Z}){\tt X} & - \omega^{2n_0}{\tt t} +  \omega^{1+2m+2n_0}{\tt Z} 
\end{array} \right) ~ ~ ({\tt t}= \omega^m \frac{t}{t_p}) ,
\elea(LFhom)
for all $\ell$. Through the dual map (\req(pp*)) and the linear isomorphism (\req(Phi)), the above model is equivalent to the homogeneous superintegrable $\tau^{(2)}$-model  with the vertical rapidity $p^*$ and boundary condition $r^*={\tt r}$, having the $L$-operator $L_\ell (t^*)$ in (\req(hsupL)) where $m, n_0, {\tt t}$ are replaced by $m^*(= m+ 2n_0), n^*_0 (= -n_0)$, and ${\tt t}^*(= \omega^{m^*} t^* t_{p^*}^{-1}= \omega^{2n_0} {\tt t})$ respectively. By Lemma \ref{lem:supspin}, the above $\tau^{(2)}$-model is equivalent to the homogeneous spin-$\frac{N-1}{2}$ XXZ chain (\req(spN-1)) via the identification (\req(ek)) for $m^*, n_0^*$. Equivalently, the $\tau^{(2)}$-face model (\req(LFhom)) is the homogeneous  XXZ chain defined by the $L$-operator
\be 
{\cal L}_F (s)   =  \left( \begin{array}{cc}
         q^{1+2m+4n_0} s K^\frac{-1}{2}   -   s^{-1} K^\frac{1}{2}   &  (q- q^{-1}) e^-    \\
        (q - q^{-1}) e^+ &   s K^\frac{1}{2} -  q^{-1-2m-4n_0}   s^{-1} K^\frac{-1}{2} 
\end{array} \right), 
\ele(xxzspF)
and the spin-$\frac{N-1}{2}$-representation (\req(spN-1)) of $K, e^\pm$ on ${\tt C}^N$, where 
\bea(ll)
{\bf e}^k:= |-m+ k \rangle \rangle ~ ~  ~ (k=0, \ldots, N-1),  & 
{\tt S}^z ({\bf e}^k) = \frac{N-1}{2} - k , 
\elea(ek*)
where ${\tt S}^{z}:= \sum_{j=1}^{N-1} \frac{ \omega^{m j }{\tt Z}_\ell^{j} }{1-\omega^{-j}}$. The second Onsager-algebra generator ${\tt H}_1$ for $\widehat{T}_F (q; p, p)$ in section \ref{ssec.CP*} becomes ${\tt H}_1 = -2 \sum_\ell {\tt S}^{z}_\ell $.  Through the isomorphism $\Theta$ in (\req(Theta)), the first Onsager-algebra generator $H_0$ in (\req(H01)) is identified with ${\tt H}_1$, $H_0 = \Theta^{-1} {\tt H}_1  \Theta $, by (\req(H0inf)).  Indeed  $\Theta^{-1}{\tt S}^{z}_\ell \Theta$ is equal to the $\ell$th local operator in $H_0$. By (\req(ttf)), the charge and boundary condition of $\tau_F^{(2)}$ and $\tau^{(2)}$ are interchanged, and we will make the identification, $({\tt r}, {\tt Q})= (Q, r)$, in later discussion. 
By the Algebraic-Bethe-Ansatz method, the $\tau_F^{(2)}$ model (\req(LFhom)) has  the pseudo-vacuum (\req(vac)) and Bethe states $\phi^\pm (= \phi^\pm ({\tt v}^\pm_1, \ldots, {\tt v}^\pm_J))$:
\bea(ll)
\phi^+ = \prod_{j=1}^{J} {\tt B}(-(\omega {\tt v}^+_j)^{-1}) \Omega^+, &(\widetilde{P}^{+}_a=0, \widetilde{P}^+_b \equiv m^*L +Q -J ); \\
\phi^- = \prod_{j=1}^{J} {\tt C}(-(\omega {\tt v}^-_j)^{-1}) \Omega^-, &(\widetilde{P}^-_a\equiv -(1+m^*)L-Q-J, \widetilde{P}^-_b=0 ),  
\elea(BetvF)
where $m^*=m+2n_0$, ${\tt A}, {\tt B}, {\tt C}, {\tt D}$ are the entries of monodromy matrix (\req(MtF)), and ${\tt v}^\pm_j$'s satisfy the Bethe equation  (\req(Bethesup)) for $\tau^{(2)}_F(t)= \tau^{(2)}(t)$ expressed by (\req(stauev)) via (\req(ttf)). 
Here we use the equivalence (\req(tf=t)) and the relation of quantum numbers in (\req(duqn)). 
The Bethe state $\phi^+, \phi^- $ in (\req(BetvF)) are with the boundary condition $r \equiv -(1+m)L-J, - m L+J$ respectively, and they are characterized as the vector with maximum or minimum $-2{\tt S}^z$ in $\Theta ({\cal E}_{{\tt F}^\pm, P^\pm_a , P^\pm_b })$. Then follows  $\Theta^{-1}(\phi^+) = \vec{v} (+, \ldots, +; 0 )$, $\Theta^{-1}(\phi^-) = \vec{v} (-, \ldots, -; 0 )$ by (\req(vec*)).  In the sector (\req(Sect)), equivalently  $P_a=\widetilde{P}_a^\pm =0, P_b= \widetilde{P}_b^\pm  =0$ in (\req(BetvF)), the Bethe states $\phi^\pm $ for ${\tt v}_j = {\tt v}^\pm_j$ correspond to $\vec{v} (\ldots, \pm , \ldots ; 0 ) \in {\cal E}_{{\tt F}, P_a, P_b}$. Indeed as before, they are the highest and lowest weight vectors of two $sl_2[z, z^{-1}]$-representations on  ${\cal E}_{{\tt F}, P_a, P_b}$, one from the $sl_2$-loop-algebra symmetry of XXZ chain using the evaluation polynomial ${\tt P}({\tt t})/N$, the other by $\rho_{0}$ in (\req(rhok')) with ${\rm e}^{{\rm i} \theta_i} (\neq {\tt t}^N_i)$ in (\req(OAev)) as the evaluation-values. The decomposition of ${\cal E}_{{\tt F}, P_a, P_b}$ in (\req(Ek')) at $k'=0$ provides the  underlying $(\stackrel{m_E}{\oplus} sl_2)$-structure for both $sl_2[z,z^{-1}]$-structures. In this situation, the Bethe states $\psi^\pm, \phi^\pm$ in (\req(Betv)), (\req(BetvF)) belong to the same $\tau^{(2)}$-eigenspace ${\cal E}_{{\tt F}, P_a, P_b}$ with $\psi^\pm=\vec{v}(\ldots,\pm, \ldots; \infty)$, $ \phi^\pm= \vec{v}(\ldots, \pm , \ldots; 0)$. Indeed, the $(\oplus sl_2)$-structures of ${\cal E}_{{\tt F}, P_a, P_b}$ at $k'=0,\infty$ of the $sl_2$-loop-algebra symmetry are identified under the correspondence, $\vec{v}(s_1, \ldots, s_{m_E}; \infty) \mapsto \vec{v}(s_1, \ldots, s_{m_E}; 0)$.
We now summary the results of this subsection as follows:
\begin{prop}\label{prop:supintfF}
$({\tt I})$.  For odd $N$ and a rapidity $p=(x_p, y_p, \mu_p)$ with $x_p= \omega^my_p, \mu_p= \omega^{n_0}$, 
the homogeneous superintegrable $\tau^{(2)}$-model $(\req(hsupL)) ~ ($ or $\tau_F^{(2)}$-model $(\req(LFhom)) )$ with the vertical rapidity $p$ is equivalent to the homogeneous spin-$\frac{N-1}{2}$ XXZ chain $(\req(xxzsp))$ $(~(\req(xxzspF)) ~resp.)$ with the $U_q(sl_2)$-representation $(\req(spN-1))$ via the basis $(\req(ek))$ $( ~(\req(ek*)) ~ resp.)$.  \par \noindent
$({\tt II})$. The Bethe states of $\tau^{(2)}$-model, $\psi^\pm$ in $(\req(Betv))$, are equal to $\vec{v} (\ldots, \pm , \ldots; \infty )$ in ${\cal E}_{{\tt F}^\pm, P^\pm_a, P^\pm_b}$ with maximum or minimum $-2S^z (=H_1)$. Through the isomorphism $\Theta$ in $(\req(Theta))$, the Bethe states of $\tau_F^{(2)}$-model, $\phi^\pm$ in $(\req(BetvF))$, corresponds to the vector in ${\cal E}_{{\tt F}^\pm, \widetilde{P}^\pm_a, \widetilde{P}^\pm_b}$ with maximum or minimum $-2{\tt S}^z (=H_0)$, $\Theta^{-1}(\phi^\pm) = \vec{v} (\ldots, \pm , \ldots; 0 )$. \par \noindent
$({\tt III})$. In the sector $(\req(Sect))$, equivalently  $P_a=P_a^\pm =\widetilde{P}_a^\pm =0, P_b= P_b^\pm = \widetilde{P}_b^\pm  =0$, the Bethe states, $\psi^\pm$ and $\Theta^{-1}(\phi^\pm) $, with ${\tt v}_j = {\tt v}^\pm_j$ in $(\req(Betv)), (\req(BetvF))$ belong to the same $\tau^{(2)}$-eigenspace ${\cal E}_{{\tt F}, P_a, P_b}$. They are the highest  or lowest weight vectors for the $sl_2[z, z^{-1}]$-representation induced from the $sl_2$-loop-algebra symmetry of XXZ chain with the evaluation polynomial ${\tt P}({\tt t})/N$ for ${\tt P}({\tt t)}$ in $(\req(sPt))$. This $sl_2$-loop-algebra symmetry and the Onsager algebra symmetry of CPM shares an unified underlying $(\oplus sl_2)$-structure of ${\cal E}_{{\tt F}, P_a, P_b}$ presented by the basis in $(\req(Ek'))$ at $k'=\infty, 0$. 
\end{prop}
{\bf Remark}. (1) The $P_a^\pm, P_b^\pm, \widetilde{P}_a^\pm, \widetilde{P}_b^\pm$ in (\req(Betv)) and (\req(BetvF)) provide the four cases in the last two constraints  in  (\req(Pab)). Indeed, the cases, $P_b^-=0$, and $\widetilde{P}_b^-=0$, correspond to those in (\req(Pab)) with $P_b \equiv 0, J +P_b \equiv (m+ 2n_0)L+Q,  mL+r$ respectively. Hence the Bethe vectors (\req(Betv)) of $\tau^{(2)}$-model derived from Algebraic-Bethe-Ansatz cover only one"half" of states, not the whole theory as previously indicated in \cite{R06F} section 3.1. The other half of states are those corresponding to Bethe states of $\tau_F^{(2)}$-model in (\req(BetvF)). \par \noindent
(2) We consider only the odd $N$ case in Proposition \ref{prop:supintfF}. However, the pseudo-vacuum discussion in (\req(ek)), (\req(ek*)) and (\req(vac)) about the ground state is valid for an arbitrary $N$ with ${\tt F}=1$, $P_a$ and $P_b$ in (\req(Betv)) or (\req(BetvF)) for $J=0$. Therefore the Bethe states' characterization in $({\tt II})$ is likely true for all $N$, hence a more complete theory about the $sl_2$-loop-algebra symmetry of ${\cal E}_{{\tt F}, P_a, P_b}$ would be expected by a general argument.

\section{Concluding Remarks \label{sec.F}}
We establish a Ising-type duality relation in $N$-state CPM as a generalization of the usual Kramers-Wannier duality of Ising model when $N=2$. The approach is based on the functional relation method in CPM. We first find the duality of $\tau^{(2)}$-model under a dual correspondence of $k', k'^{-1}$- rapidities and quantum spaces for the temperature-like parameter  $k'$, under the constraint of interchanging the $\ZZ_N$-charge and skewed boundary condition. Then the duality of CPM follows from the dual connection between Boltzmann weights at $k'$ and $k'^{-1}$. The method is carried out by calculations in Fourier transform about the transfer matrix, as the duality discussion in \cite{B89} on a special superintegrable case  about the vertical interfacial tension of CPM. We can incorporate this duality into the CPM and $\tau^{(2)}$-model over the dual lattice, as well as that of the face $\tau^{(2)}$-model. The duality in this work not only establishes a complete theory about the duality symmetry of a general CPM, but also provides a useful means for gaining insights on the structure, generally not available within the superintegrable case alone. In the homogeneous superintegrable case, the duality symmetry fits nicely with the Onsager-algebra and $sl_2$-loop-algebra symmetry about the degeneracy of $\tau^{(2)}$-model. In particular, the analysis of Bethe states leads to the understanding of eigenvectors in CPM in the sector (\req(Sect)) through the Algebraic-Bethe-Ansatz method for the odd $N$ case. The approach apparently works also in the general situation as suggested in Remark (2) of Proposition \ref{prop:supintfF}. A more complete theory about eigenvectors of CPM for all sectors and an arbitrary  $N$ is now under consideration along this line. In view of the fundamental importance of Kramers-Wannier duality in the study of Ising model, regardless of the complicated nature of techniques in CPM, further development on the duality found in the present work would be expected, especially concerning the structure at the critical $k'= \pm 1$ case. This problem is out of the scope of this paper and will be considered elsewhere.

\section*{Acknowledgements}
The author is pleased to thank Professor T. Mabuchi for hospitality in October 2008 at the Department of Mathematics, Osaka University, Japan, where part of this work was carried out.

\end{document}